\documentclass[%
 reprint,
 amsmath,amssymb,
 aps,
]{revtex4-1}

\usepackage{graphicx}
\usepackage{dcolumn}
\usepackage{bm}
\usepackage[space]{grffile}
\usepackage{float}

\usepackage{amsmath}

\begin{document}

\preprint{APS/123-QED}

\title{Increased persistence via asynchrony in oscillating ecological populations with long-range interaction}


\author{Anubhav Gupta}
\homepage{anubhav.iiser@gmail.com}
\affiliation{
Department of Mathematics and Statistics\\ Indian Institute of Science Education and Research Kolkata, Mohanpur 741 246, West Bengal, India
}%

\author{Tanmoy Banerjee}
\homepage{tbanerjee@phys.buruniv.ac.in}
\affiliation{
Chaos and Complex Systems Research Laboratory, Department of Physics, University of Burdwan, Burdwan 713 104, West Bengal, India
}%

\author{Partha Sharathi Dutta}
\homepage{Corresponding author: parthasharathi@iitrpr.ac.in}
\affiliation{
Department of Mathematics, Indian Institute of Technology Ropar, Rupnagar 140 001, Punjab, India
}%

\date{\today}

\begin{abstract}
Understanding the influence of structure of dispersal network on the
species persistence and modeling a much realistic species dispersal in
nature are two central issues in spatial ecology. A realistic
dispersal structure which favors the persistence of interacting
ecological systems has been studied in [Holland \& Hastings, Nature,
  456:792--795 (2008)], where it is shown that a {\it randomization}
of the structure of dispersal network in a metapopulation model of
prey and predator increases the species persistence via clustering,
prolonged transient dynamics, and amplitudes of population
fluctuations. In this paper, by contrast, we show that a {\it
  deterministic network} topology in a metapopulation can also favor
asynchrony and prolonged transient dynamics if species dispersal obeys
a long-range interaction governed by a distance-dependent
power-law. To explore the effects of power-law coupling, we take a
realistic ecological model, namely the Rosenzweig-MacArthur model in
each patch (node) of the network of oscillators, and show that the
coupled system is driven from synchrony to asynchrony with an increase
in the power-law exponent. Moreover, to understand the relationship
between species persistence and variations in power-law exponent, we
compute correlation coefficient to characterize cluster formation,
synchrony order parameter and median predator amplitude. We further show
that smaller metapopulations with less number of patches are more
vulnerable to extinction as compared to larger metapopulations with
higher number of patches.  We believe that the present work improves
our understanding of the interconnection between the random network
and deterministic network in theoretical ecology.

\end{abstract}

\maketitle


\section{\label{sec:level1}Introduction}

What determines the persistence of interacting spatially separated
sub-populations?  Or in other words, how long will these interacting
sub-populations survive in a given time frame?  This is indeed an
important question in landscape ecology, where spatially separated
sub-populations or patches of the same species create metapopulation
structure via dispersal network \cite{Han98:Nature}.  Metapopulations
are generally dynamic in nature \cite{ilmari2003long} with frequent
local extinctions, and dispersal within patches to allow
recolonization, thereby preventing global extinction
\cite{johst2002metapopulation}.  Metapopulation dynamics can be used
to understand various ecological processes, including a variety of
ecological and evolutionary dynamics that includes population size
\cite{gyllenberg1992single}, species persistence
\cite{roy2005temporal}, spatial distribution
\cite{roy2008generalizing}, epidemic spread
\cite{mccallum2002disease}, gene flow \cite{sultan2002metapopulation}
and local adaptation \cite{joshi2001local}.  It is necessary to
understand the consequences of dispersal network structure because
depending on it the metapopulation dynamics evolve
\cite{hansson1991dispersal}.

Dispersal is crucial for the survival of metapopulation as species
that are connected through dispersal are less prone to extinction than
that of disconnected ones \cite{holyoak1996persistence}. On the
contrary, there is a very high chance of global extinction if the
patches are synchronized due to dispersal
\cite{heino1997synchronous}. Dispersal can be called a ``double-edged
sword'' \cite{hudson1999moran} because it can facilitates persistence
at the local scale, but also leads to synchrony, and hence an
increased probability of extinction. Therefore, there should be
sufficient asynchrony in the patches to ensure species persistence
\cite{hanski1995metapopulation,ellner2001habitat}.  During asynchrony,
as the species populations among the patches are different, so species
can move from patches with higher abundance to connected patches
thereby supporting the destination patches.

There has been a large number of studies including experimental
\cite{dey2006stability, ringsby2002asynchronous} and theoretical
\cite{holland2008strong} on the ways to introduce asynchrony in
metapopulation.  Most significantly, Holland \& Hastings
\cite{holland2008strong} have shown the role of dispersal network
heterogeneity in species persistence by inducing asynchrony in the
system. They have instigated asynchrony among the sub-populations by
varying the network topology from regular to random and demonstrated
that in general heterogeneous networks have longer periods of
asynchronous dynamics, leading typically to lower amplitude
fluctuations in population abundances.

The main question we address here is: Is there some other way to
induce asynchrony and hence increased persistence in the system
without using a heterogeneous network topology?  Incorporating
long-range interactions obeying distance-dependent power-law between
the habitat patches we probe this question.  In contrast to
\cite{holland2008strong}, here we have considered the role of distance
between patches in the model with a regular network topology.

The motivation behind this is that in nature species are not likely to
move to all the patches equally. Generally, species dispersal is
dependent on the distance between habitat patches. Dispersal can be
classified into ``short-distance dispersal'' (SDD) and ``long-distance
dispersal'' (LDD).  In nature, both SDD and LDD are prevalent with
large intensity of short dispersals as compared to long
dispersals. Long-distance dispersers greatly suffer from dispersal
mortality as in a large ecological network not all the patches are
likely to be accessible from a particular patch \cite{BoBe09:Oikos}.
Hence, the density of species moving to furthest patches decreases
with increasing distance between patches, which can be seen in species
like amphibians \cite{alex2005dispersal}, butterflies
\cite{baguette2003long}, mites \cite{BoBe09:Oikos}, etc.  Although LDD
events are typically rare, they are crucial to metapopulation survival
\cite{bohrer2005effects, trakhtenbrot2005importance}.

Most metapopulation models consider only SDD while underestimating LDD
\cite{hanski1999metapopulation,baguette2003long}. This leads to
falsely estimating the scale of a metapopulation effect by reducing
the extent of global dynamics to regional dynamics.  Therefore, it is
necessary to correctly incorporate LDD together with SDD into
studies. One can incorporate both SDD and LDD into spatial ecological
models by using distance-dependent power-law coupling. Previous
empirical studies have shown that inverse power-law gave a better fit
to empirical dispersal data
\cite{hill1996effects,thomas1997butterfly,baguette2000population}.
Power-law dispersal is more general and universal coupling scheme
which is motivated by many real-world systems. In the long-range
coupling, each patch is connected to all the other patches with an
effective dispersal strength according to power-law whose interaction
strength is governed by an exponent (denoted by $s$).  Studies have
been performed using long-range interaction obeying power-law coupling
in ferromagnetic spin models \cite{PhysRevB.54.R12661}, hydrodynamic
interaction of active particles
\cite{PhysRevLett.106.058104,C0SM01121E}, ecological networks
\cite{PhysRevE.94.032206}, biological networks
\cite{PhysRevLett.74.3297}, etc.  The power-law exponent $s$
represents how likely the species is to travel to a further habitat. A
higher value of $s$ suggests that the likelihood of a species to
travel to a further habitat is less as compared to a lower value of
$s$.

Here we analyze an ecological network of habitat patches whose local
dynamics are governed by the Rosenzweig-MacArthur model \cite{RoMa63}
and the dispersal between the patches is governed by a long-range
interaction obeying a distance-dependent power-law. The effective
strength of long-range interaction between habitat patches decreases
with increasing power-law exponent. In this paper, we show the
dramatic transformation in the spatiotemporal dynamics when species
are connected by power-law. We compute numerical measures like cluster
distribution, interpatch synchrony, predator amplitude and transient
time to understand species persistence. Our key finding is the
asynchronicity that is generated, surprisingly, even in a regular
network by incorporating power-law dispersal.  Furthermore, by varying
the network size we find that metapopulations with larger number of
patches are more persistence (hence less prone to extinction) in
comparison with metapopulations with less number of patches.

We organize the paper as follows: First, we discuss the structure of
ecological network and the model used to govern the local dynamics,
along with the coupling used to model dispersal dynamics between
patches in Sec.~\ref{II}. Then, we present the effects of power-law on
the spatiotemporal dynamics of the model in Sec.~\ref{III} by
calculating several dynamical measures. Here, we also show the effect
of network size in species persistence. Finally, in Sec.~\ref{IV} we
discuss the importance of our findings.

\begin{figure*}
\centering
\includegraphics[width=0.3\textwidth]{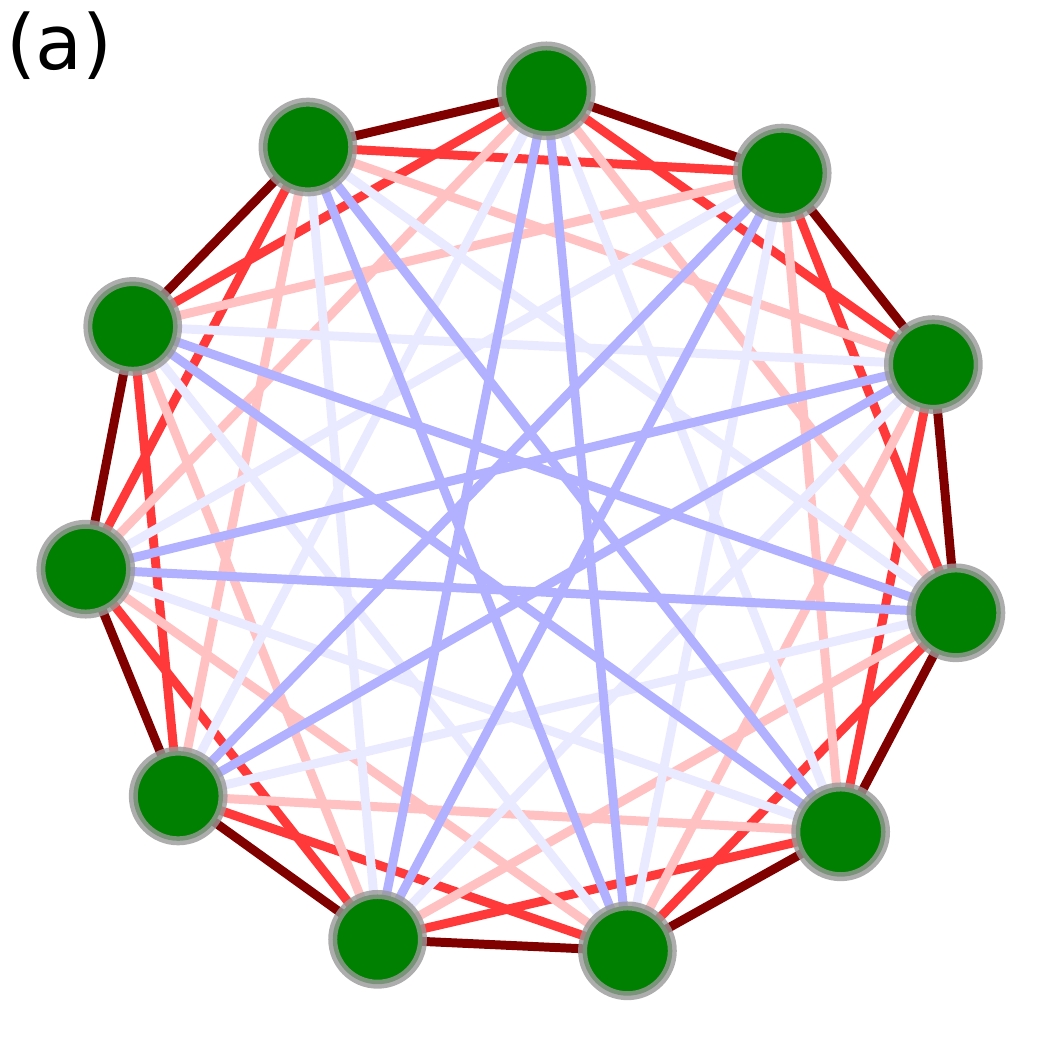}\hspace{0.2in}
\includegraphics[width=0.3\textwidth]{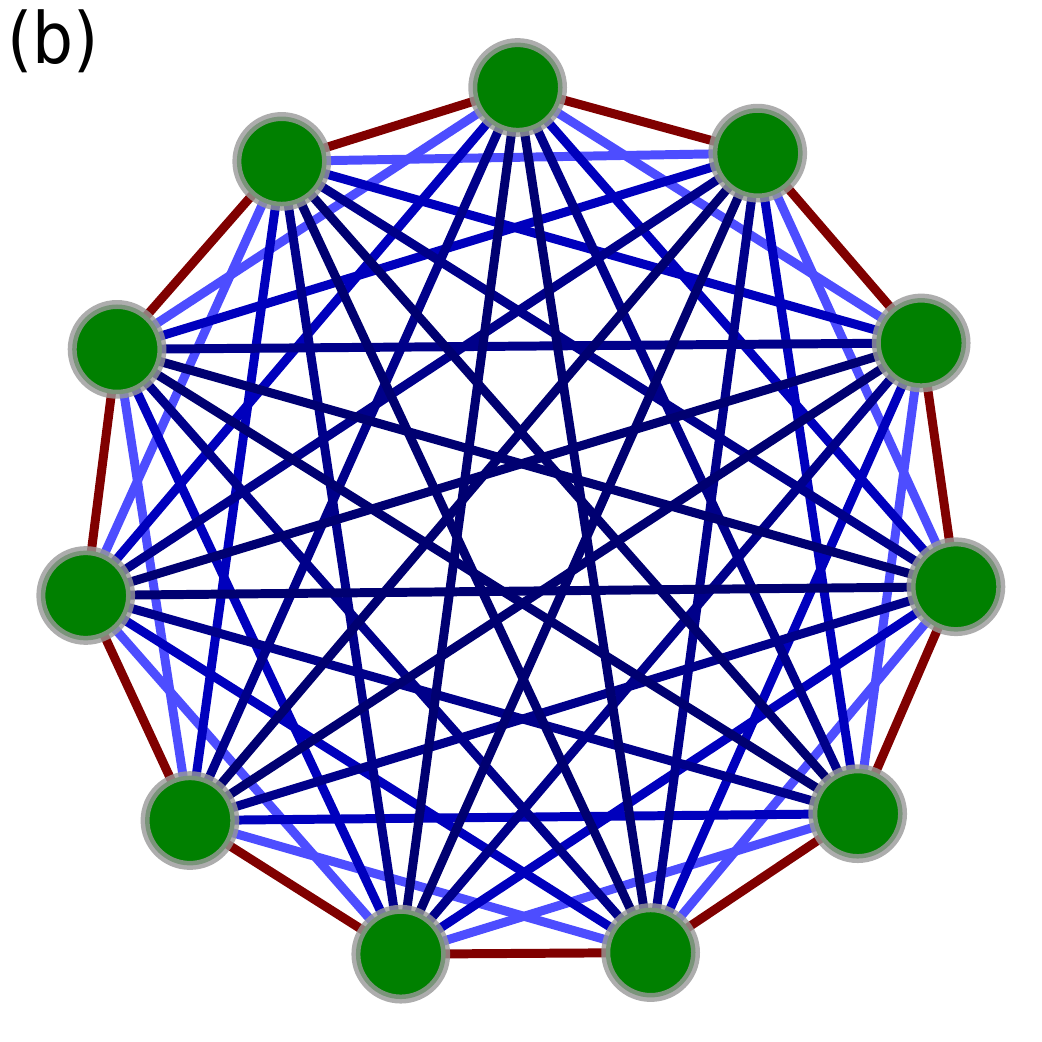}\hspace{0.2in}
\includegraphics[width=0.135\textwidth]{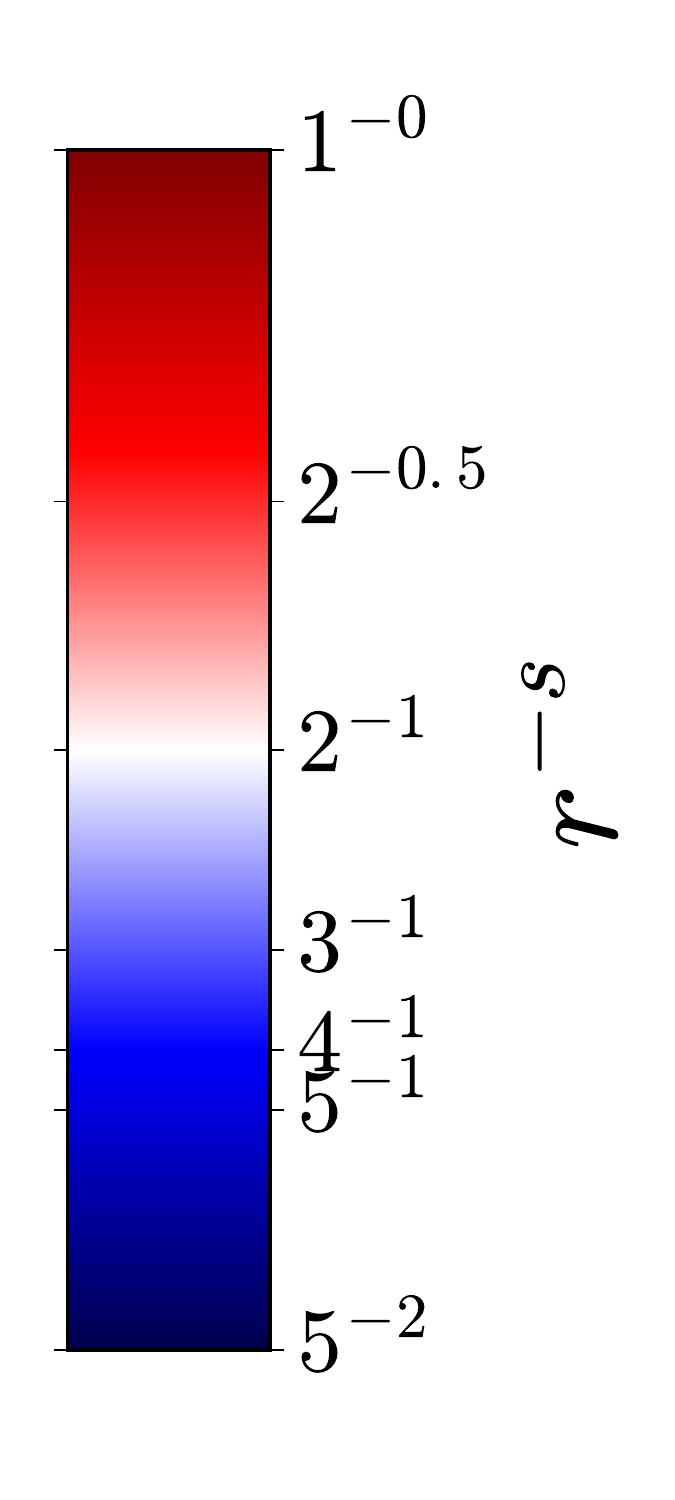}
\caption{(Color online) All patches are connected to all patches in
  the network, and dispersal is bi-directional following the
  long-range interaction as modeled in Eq. (1). Color of the edges
  represents power-law strength ($r^{-s}$) between patches with: (a)
  $s = 0.5$ and (b) $s = 1.5$. Higher $s$ suggests that the species is
  less likely to move to further patches. Each spatially separated
  patches are marked with filled circles.  Here we consider an
  ecological network with 11 patches.}\label{fig1}
\end{figure*}

\section{The Network Model}\label{II}

We consider an $n$-patch/node prey-predator system where uncoupled
dynamics in each patch are governed by a non-dimensional form of the
well known Rosenzweig-MacArthur model \cite{RoMa63}.  A schematic
picture of the ecological network is depicted in Fig.~\ref{fig1}.  In
the network, we assume that dynamics within each patch are identical
and the dispersal dynamics of prey-predator are modeled as follows:
\begin{subequations}\label{eq1}
\begin{align}
\frac{dh_i}{dt} &= h_i (1-\theta h_i) - \frac{p_i h _i}{1+h_i}\nonumber \\
&\;\;\;\; + d_h \left(\frac{1}{\xi(s)}\sum_{r=1}^{m}\frac{h_{i-r}+h_{i+r}}{r^s} - h_i\right),\\
\frac{dp_i}{dt} &= \frac{\phi p_i h_i}{1+h_i} - \eta p_i + d_p \left(\frac{1}{\xi(s)}\sum_{r=1}^{m}\frac{p_{i-r}+p_{i+r}}{r^s} - p_i\right),
\end{align}
\end{subequations}
where $i = 1, 2, \dots, n$ is the patch index, the prey and predator
density are represented by $h_i$ and $p_i$, respectively for the
$i$-th patch with all indices are taken modulo $n$ (the total number
of patches in the network).  The local (uncoupled) dynamics in each
patch are governed by the following parameters: $\theta$ is the
strength of prey self-regulation, $\phi$ is the predator conversion
efficiency, and $\eta$ is the predator mortality rate. The spatial
dynamics are governed by: dispersal rate $d_h \text{ and } d_p$ for
prey and predator, respectively, the dispersal range $m$, and
$\xi(s)=2\sum_{r=1}^{m} r^{-s}$ is the normalization constant. The
interaction between the neighboring patches follows a dispersal rate
whose intensity decays with the distance $r$ between patches as
inverse power law $r^{-s}$.  Here, $s \geq 0$ is the power-law
strength \cite{PhysRevLett.74.3297} and $r$ $(r = 1, \dots, m)$ is the
distance between $i$-th and $j$-th patches, which is defined to be the
minimum number of edges required to move from the $i$-th patch to the
$j$-th patch along the arc with $m = (n-1)/2$ for odd number of
patches.

The significance of this model (\ref{eq1}) of ecological network with
long-range coupling is that one can take into account both SDD and LDD
along with their intensities. Also it is possible to control the
effective LDD by varying the parameter $s$ only.  For instance, when
$s = 0$, a species is equally likely to move into any patch resulting
in the equal intensities of SDD and LDD, whereas when $s>0$, a species
is less likely to move to a farther patch as compared to a nearby
patch resulting in less intensity of LDD as compared to SDD.  However,
this form of connectivity is different from the usual ecological
network models with a fixed dispersal rate
\cite{holland2008strong,GoHa08,GoHa11,hastings2001transient,wall2013synchronization}.

\begin{figure*}
\centering
\includegraphics[width=0.78\textwidth]{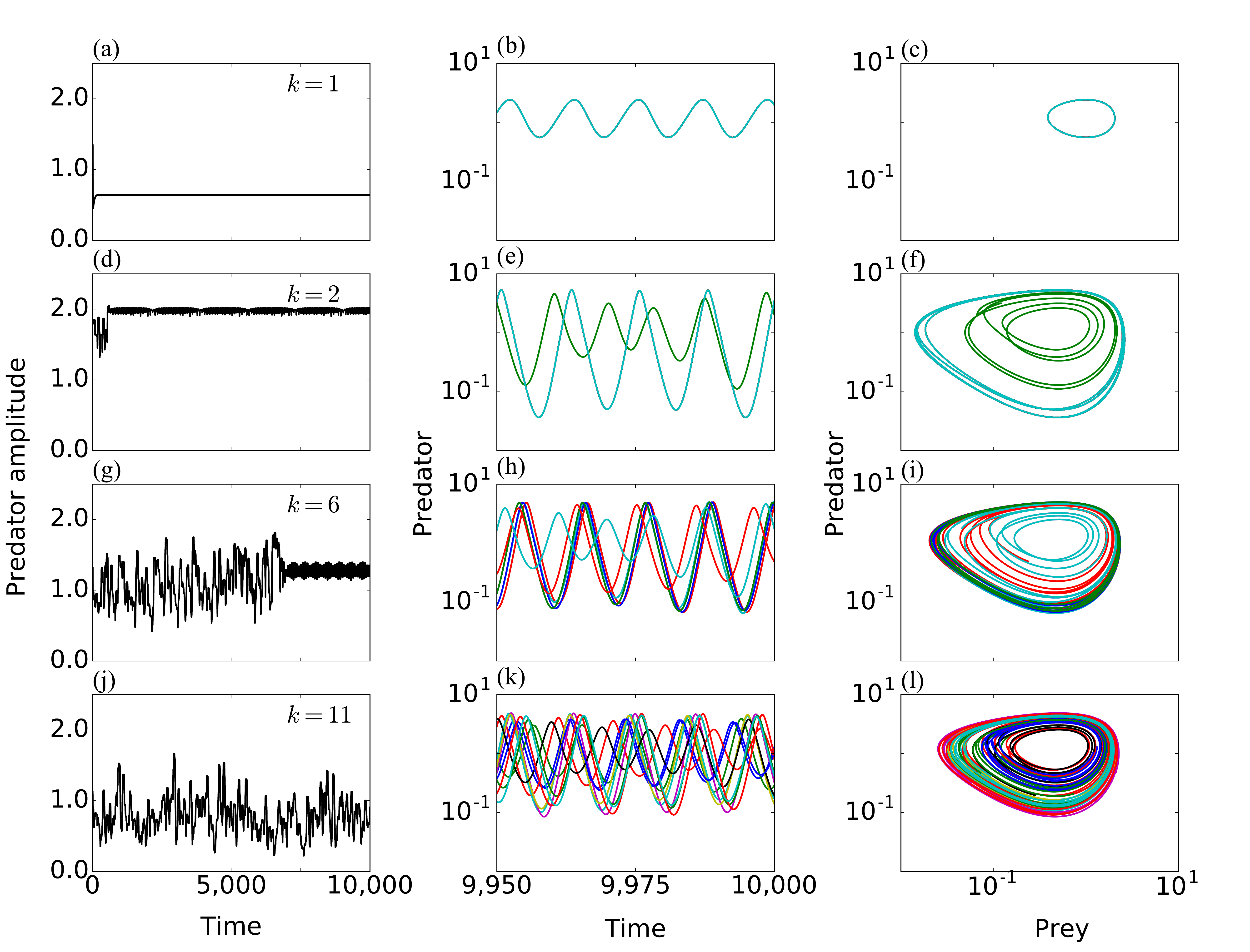}
\caption{(Color online) Time series of ((a), (d), (g), (j)) total
  predator amplitude and ((b), (e), (h), (k)) predator species, and
  ((c), (f), (i), (l)) phase portrait of predator vs prey for
  different cluster solutions.  Local dynamics are governed by strong
  prey self-regulation $\theta = 0.3$ and predator mortality rate
  $\eta = 1$ comparable to prey birth rates.  Dispersal rates are:
  $d_h = 2^{-5}$ and $d_p = 2^{-6}$. Other parameters are: ((a), (b),
  (c)) $s = 0, \phi = 2, k = 1$; ((d), (e), (f)) $s = 0, \phi = 3, k =
  2$; ((g), (h), (i)) $s = 1, \phi = 3, k = 6$; and ((j), (k), (l)) $s
  = 2, \phi = 3, k = 11$.}\label{fig2}
\end{figure*}

\section{Results}\label{III}

We explore the spatiotemporal dynamics of the coupled
Rosenzweig-MacArthur model (\ref{eq1}) with variations in the
parameter $s$.  We always fix the dispersal range to be $m = (n-1)/2$
(i.e., we consider a globally coupled network) and vary $s$.  In other
words, all the patches are accessible by a disperser from any patch,
however the dispersal density between them will depend on the distance
between the patches and hence on the value of
$s$. Therefore, we start with a globally coupled network ($s = 0$) and
vary the $s$, effectively reducing the strength of long-range
interactions.

Our main goal is to show that asynchrony can be introduced into the
system by varying the power-law exponent $s$.  We investigate the
effect of $s$ on species persistence and calculate measures like
cluster identification, parameter of synchrony, median predator amplitude
and mean transient fraction to justify our key findings. We also study
how predator dispersal $d_p$ and $\phi$ affect the results.  Before we
proceed, let us discuss the integration scheme we used
to solve Eqs.~(\ref{eq1}).  We perform the numerical integration
using backward-differentiation formula method in CVODE
\cite{cohen1996cvode,holland2008strong}.  Prey initial conditions are
independently and identically distributed, with $\log_{10}(h_i(0))$
uniformly distributed on the interval
$(-5,1+\log_{10}\hat{h})$. Similarly, predator initial conditions are
independently and identically distributed, with $\log_{10}(p_i(0))$
uniformly distributed on the interval $(-5,1+\log_{10}\hat{p})$, where,
$\hat{h} = \frac{\eta}{\phi-\eta}$ and $\hat{p} = (1+\hat{h})(1-\theta
\hat{h})$ \cite{holland2008strong}.

\begin{figure}
\centering
\includegraphics[width=0.47\textwidth]{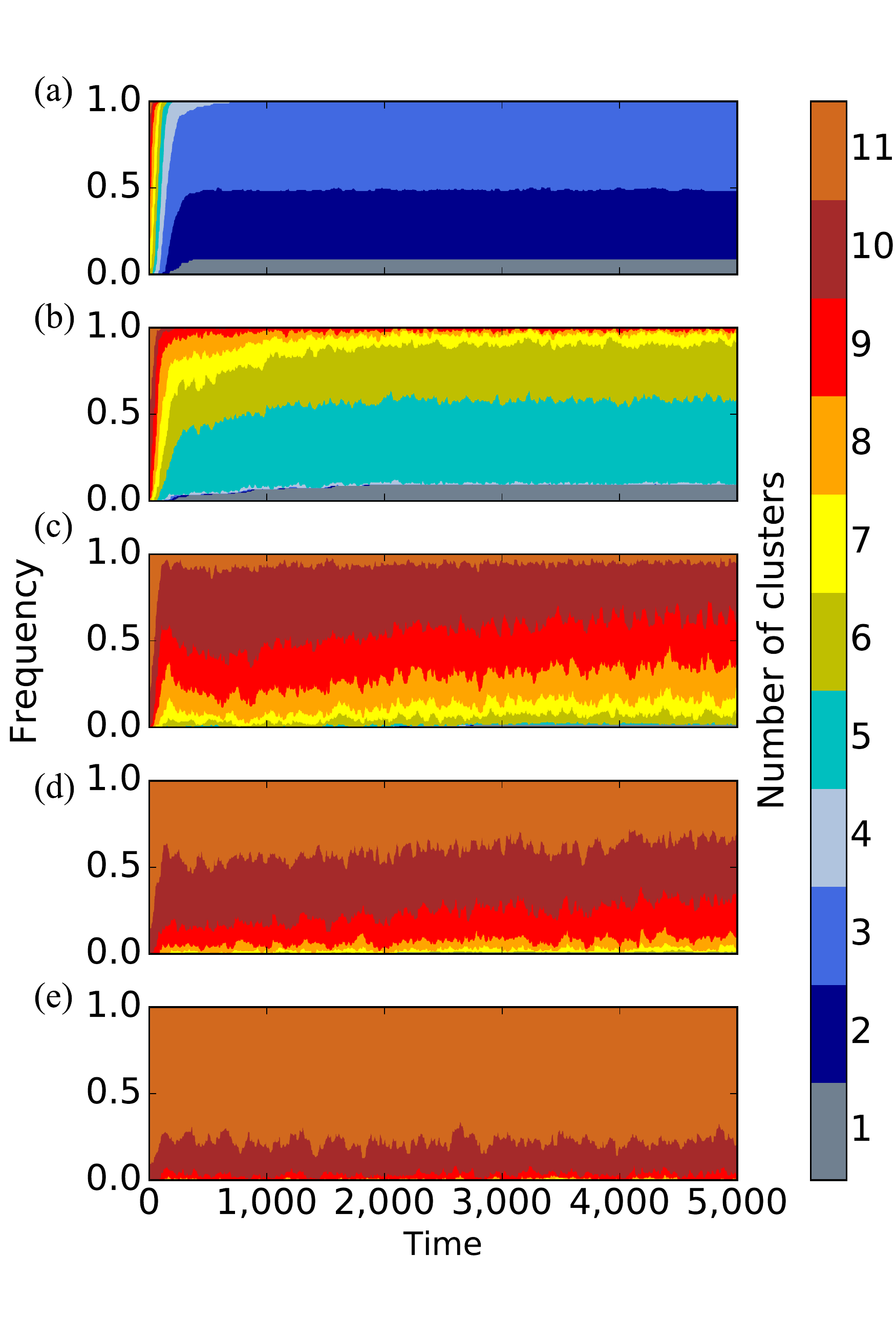}\vspace{-0.28in}
\caption{(Color online) Distribution of cluster states for different
  values of $s$: (a) $s = 0$; (b) $s = 0.3$; (c) $s = 0.6$; (d) $s =
  0.9$; and (e) $s = 1.2$. The local dynamics are governed by weak
  predation $(\phi = 2.7)$, strong prey self-regulation $(\theta =
  0.3)$ and predator mortality rate $(\eta = 1)$ comparable to prey
  birth rates.  The spatial parameters are $d_h = 2^{-5}$ and $d_p =
  2^{-6}$. Number of clusters increases with increasing values of $s$.
  The above results in each sub-figures correspond to $100$ numerical
  simulations carried out independently.}\label{fig3}
\end{figure}

\subsection{Cluster analysis}

It is interesting to observe that Eqs.~(\ref{eq1}) can show a variety
of spatial dynamics, ranging from global synchrony to global
asynchrony. Cluster analysis is used to study the synchronous dynamics
of the system. To compare the dynamics between a pair of patches
$(i,j)$, linear correlation coefficient $\rho_{ij}$ of prey time
series at time $t$ is calculated as follows:
\begin{equation*}
\rho_{ij} = \frac{<h_i h_j> - <h_i><h_j>}{\sqrt{<h_i^2>-<h_i>^2}
    \sqrt{<h_j^2>-<h_j>^2}},
\end{equation*}
where, $<\dots>$ is average over the window $[t, t + 4\overline{T}]$,
with $\overline{T}$ as the mean period of a predator-prey cycle for a
given simulation, averaged over all patches. We calculate the linear
correlation coefficient $\rho_{ij}$ at each unit time.

Two patches would behave identically with time if they have $\rho_{ij}
> 0.999$. So, we segregate the patches into clusters and define a
cluster to be a set of patches with $\rho_{ij} > 0.999$. We call a
$k$-cluster to be the maximum number of clusters. In this way, a
$k$-cluster solution could be formed where the $n$ patches could be
assigned to $k~(\leq n)$ clusters of patches with identical dynamics.
We compute the frequency of $k$-cluster solution with time, where the
frequency at time $t$ is defined as:
\begin{equation*}
\mbox{Frequency of $k$-cluster solution} = \frac{\mbox{No. of $\leq
    k$-clusters }}{\mbox{No. of simulations}}
\end{equation*}

Apart from global synchrony (we may call one cluster solution) and
global asynchrony (n-cluster solution), Eqs.~(\ref{eq1}) can result in
intermediate solutions between two and ($n-1$)-clusters of synchronous
patches with variations in $s$. These $k$-cluster solutions may vary
over time thereby converging to a single $k$-cluster solution with
higher asynchrony in the initial transient phase. We observe different
kinds of cluster behaviors which can be clearly seen in
Fig.~\ref{fig2}.  It is important to keep in mind that there might be
several $k$-cluster solutions ({\em c.f.}  Fig.~\ref{fig3}) for $2
\leq k \leq n$ even for same parameter values.  This is due to the
fact that the network being a higher dimensional system it has
multiple steady states, thereby making it necessary to carry out a
large ensemble of simulations with a collection of different initial
conditions.  Figure~\ref{fig3} shows the variation in solutions for
different power-law exponent $s$.  Surprisingly we see that the system
is driven from synchrony to asynchrony by varying the long-range
interaction through $s$ even with a regular network topology.  We can
say that the species is more persistent when $s$ is increased, as
asynchronous patches are less prone to extinction than synchronous
patches because synchronous patches are more vulnerable to
environmental perturbation as minimum level of all sub-populations
occur simultaneously at some time
\cite{holyoak1996persistence}. Asynchrony also results in rescue
effect as the species can move from a patch with higher population to
a patch with lower population thereby recolonizing the population.

\begin{figure}
\centering
\includegraphics[width=0.47\textwidth]{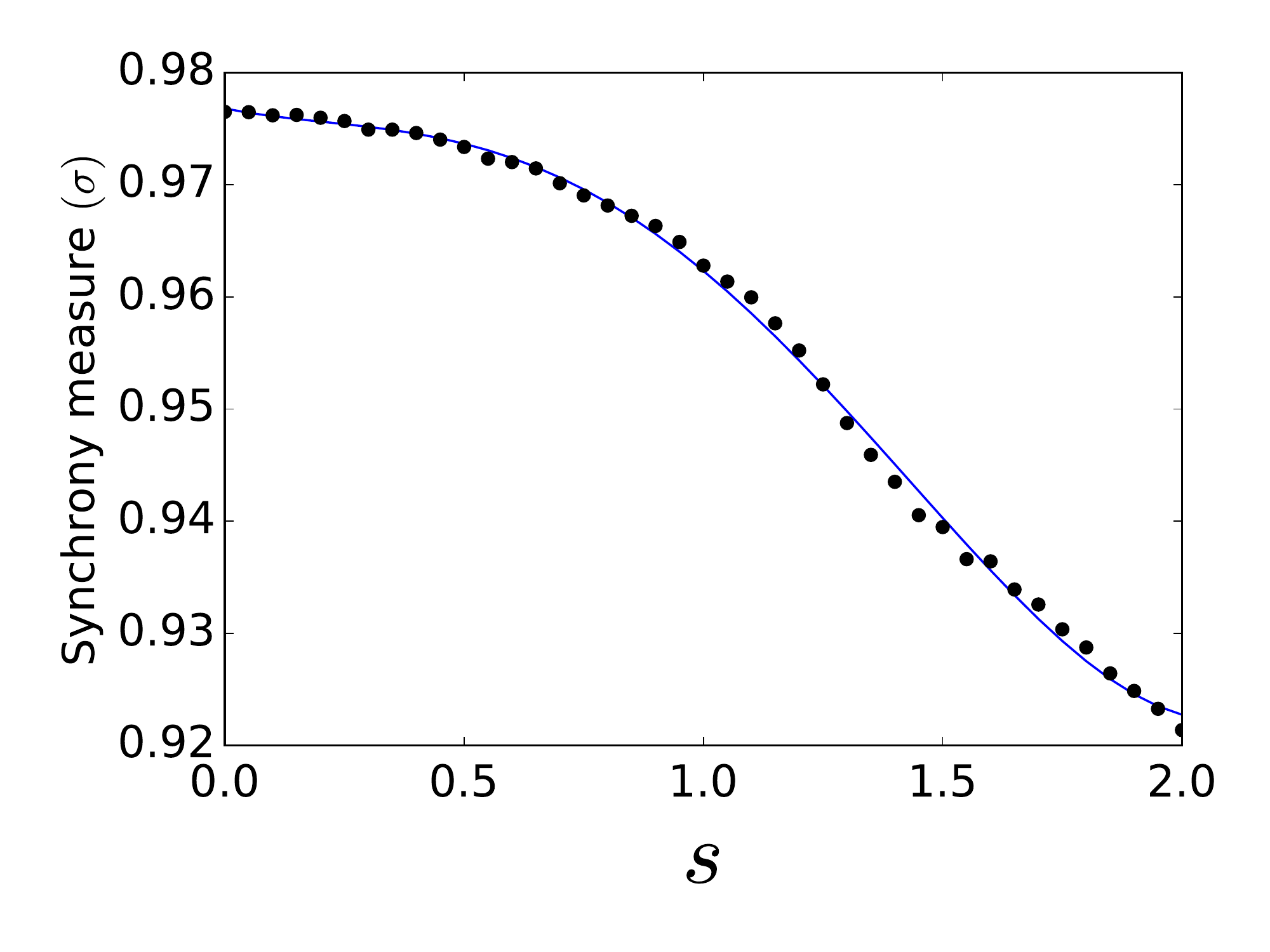}\vspace{-0.22in}
\caption{(Color online) Synchrony order parameter ($\sigma$) with
  variations in $s$. The other parameters are $d_h = 2^{-10}$, $d_p =
  2^{-10}$, $\phi = 2$, $\eta = 1$, and $\theta = 0.3$. Network size
  is 11. Points are fitted by a polynomial of degree 4 (blue curve).
  The synchrony order parameter ($\sigma$) decreases with increasing
  $s$.  The above result corresponds to $100$ numerical simulations
  carried out independently.}\label{fig4}
\end{figure}

\subsection{Measure of interpatch synchrony}

We also measure the amount of change in interpatch synchrony
with variations in $s$ (see Fig.~\ref{fig4}).  To quantify
the effect of change in long-range interaction in system dynamics, we
apply a synchrony order parameter (denoted by $\sigma$)
\cite{komin2011synchronization} for a time period of 10000 and vary
$s$, where:
\begin{equation*}
\sigma = \sqrt{1- \Bigg \langle \frac{\sum_{i=1}^n[h_i(t)-
      \overline{h}(t)]^2}{\sum_{i=1}^n h_i(t)^2} \Bigg \rangle},
\end{equation*}
with $\overline{h}(t) = \frac{1}{n}\sum_{i=1}^n h_i(t)$ and $\Big
\langle \dots \Big \rangle$ denotes the average over mentioned time.
The value of the parameter $\sigma$ varies between 0 and 1. It is
equal to 0 when there is no synchrony and 1 if the patches are
perfectly synchronized.  The value is in between 0 and 1 when the
patches are partially synchronized.  We find that with an increase in
$s$, the synchrony order parameter decreases.  As shown in
Fig.~\ref{fig4}, the interpatch synchrony decreases with increasing
$s$ and hence species persistence increases as we increase $s$.  We
fit a curve with the synchrony parameter values $\sigma$ for visual
guidance.

\subsection{Total predator amplitude}

As the species in a patch is oscillating with time there will be
instances of time when the population would be at its minimum. The
species would be at risk if the minimum densities occur at the same
time instance for all the patches.  However, the chances of extinction
can be reduced if the minimum populations corresponding to different
patches occur at considerably different time.  In other words, we
could say that a population would have higher extinction risk if the
fluctuations are high. To quantify these fluctuations we calculate
total predator amplitude, following \cite{holland2008strong} which is
defined as:
\begin{equation*}
\mbox{Total predator amplitude} = \log_{10}\left(\frac{max(\sum_{i=1}^n
  p_i)}{min(\sum_{i=1}^n p_i)}\right),
\end{equation*}
over the window of interest $4 \overline{T}$.  The way the total
predator amplitude is defined, we could say its value would be high
when the populations are synchronized while the value would be low if
the populations are asynchronous.  Figures~\ref{fig2}(a), (d), (g) and
(j) shows the amplitude fluctuations with time for different parameter
values. The amplitude fluctuations would depend on the number of
clusters as higher number of clusters would correspond to lower total
predator amplitude which can be clearly seen in Fig.~\ref{fig2}.

\begin{figure}
\centering \includegraphics[width=0.47\textwidth]{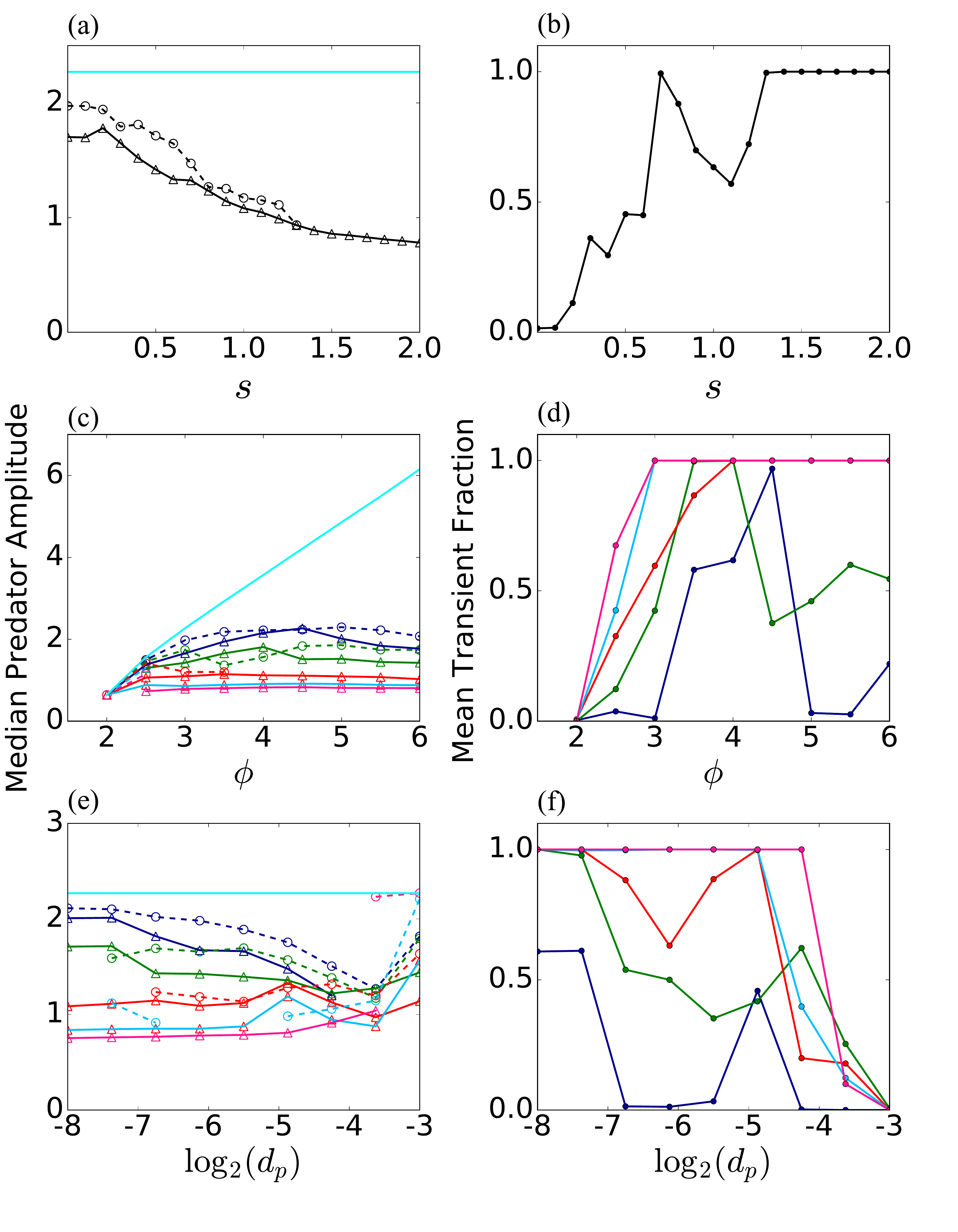}\vspace{-0.2in}
\caption{(Color online) Predator amplitude and transient duration:
  ((a), (c), (e)) Median total predator amplitude during transient
  (triangles, solid lines) and asymptotic (open circles, dashed lines)
  solution phases for a regular network; ((b), (d), (f)) mean
  transient fraction for different parameters. The cyan line in ((a),
  (c), (e)) corresponds to global synchronous solution for the chosen
  parameter values: ((a), (b)) $\phi = 3.0, \eta = 1, \theta = 0.3,
  d_h = 2^{-5}, d_p = 2^{-6}$; ((c), (d)) $\eta = 1, \theta = 0.3, d_h
  = 2^{-5}, d_p = 2^{-6}$; and ((e), (f)) $\phi = 3.0, \eta = 1,
  \theta = 0.3, d_h = 2^{-5}$. The color coding's are as follows: $s =
  0.0$ (dark blue); $s = 0.5$ (green); $s = 1.0$ (red); $s = 1.5$
  (light blue); and $s = 2.0$ (pink).  The above result corresponds to
  150 numerical simulations carried out independently.}\label{fig5}
\end{figure}

We divided the time series into two phases: transient and asymptotic
phases, where transient phase is defined as the time before all the
patches display constant or periodic phase evolution. The remaining
time is defined to be the asymptotic phase.  There are instances when
the populations would have never reached the asymptotic phase for the
given duration ($t=10000$).  We used an automated algorithm as described
in Ref.~\cite{holland2008strong} to estimate transient time from
numerically integrated solutions of Eqs.~(\ref{eq1}).

The main idea behind dividing the time into two phases rather than
only considering the asymptotic phase is because studies
\cite{hastings2004transients,hastings2001transient} have shown the
necessity to consider both of them.  In most of the studies it is
assumed that ecological systems that are observed in nature correspond
in some way to stable equilibria \cite{hastings2004transients}. This
is not the case because for example, if we want to study the
persistence of plankton species \cite{huisman1999biodiversity}, the
seasonality effectively reduces the relevant timescale to less than
one year especially in temperate lakes, then we have to consider the
model behavior within a singe season and then restart the model the
next season, rather than looking at the long term result of the
model. Also in the case of adaptive management of renewable resources,
it is necessary to understand the short-term responses than the
long-term outcomes \cite{hastings2004transients}.

As we have divided time into two phases: transient and asymptotic, we
separate the total predator amplitude into two time series and then we
take the median of these time series which we call as median predator
amplitude (MPA) for transient and asymptotic phases.
Figure~\ref{fig5} depicts MPA and transient time for different
parameters. It is important to note that here MPA is taken to be the
average of 150 ensemble of simulations because different initial
conditions might lead to different spatiotemporal dynamics. As MPA
tells us how much the total population is fluctuating, so a higher MPA
would correspond to a lower species persistence and vice
versa. Figures.~\ref{fig5}(a) and \ref{fig5}(b) represent variations
in MPA and transient time respectively with $s$. Interestingly, near
$s = 0$ (i.e., when dispersal is independent of distance) we observe
both transient and asymptotic MPA to have a higher amplitude closer to
that of a global synchronous solution which has a high extinction risk
because populations in all the patches have identical dynamics and any
perturbation when the population is very low might lead to global
extinction. This global extinction can be avoided if populations in
all the patches do not behave exactly the same with time (i.e., they
are asynchronous). Importantly, with increasing $s$, there is a
decrease in amplitudes leading to higher species persistence.
Surprisingly, MPA at transient phase is lower than MPA at asymptotic
phase which suggests that the species have higher extinction risk
during asymptotic phase than at the transient phase. The reason behind
missing data point for MPA in asymptotic phase is that for higher
values of $s$ the system never reaches asymptotic phase in 10000 units
of time.  For lower values of $s$, we see very low transient time in
Fig.~\ref{fig5}(b) and with increasing $s$, mean transient fraction
saturates to 1.  These higher transient time corresponds to
asynchronous solutions with large number of clusters as also depicted
in Fig.~\ref{fig3}.

We also investigate the effect of predator efficiency $\phi$ on MPA
and transient time.  Therefore, we plotted MPA and transient time for
different values of $s$ with respect to $\phi$ in Figs.~\ref{fig5}(c)
and \ref{fig5}(d).  One can conclude that by varying predator
efficiency, $\phi$, the system can be driven from low to high MPA. For
higher predator efficiencies, the global synchronous solution has
higher order of magnitudes thereby making global extinction very much
likely. An important result is, for any value of $\phi$, MPA decreases
and transient fraction increases or saturates to one with $s$ from
which it can be concluded that the system is driven to higher
persistence. However, for a very low value of $\phi$ (i.e., $\phi=2$),
the system has globally synchronous solution for any value of $s$,
thereby having a very high chance of global extinction. From
Fig.~\ref{fig5}, one might say that the system with higher values of
$s$ spend much more time on lower-amplitude transient solutions.

Another important parameter to consider in this study is the predator
dispersal rate $d_p$. Increasing the predator dispersal from a low
value, the system moves from longer to shorter transients.  For almost
all dispersal rates in Figs.~\ref{fig5}(e) and \ref{fig5}(f), the
system moves from high to low MPA and shorter to longer transients
when $s$ has increased.  For $s = 0$, there is almost no transient
resulting in no transient MPA for some predator dispersal rate, as
shown in Figs.~\ref{fig5}(e) and \ref{fig5}(f).  The transient fraction
saturates to one for higher values of $s$.  However, for high value of
dispersal rate, we observe global synchronous solution irrespective of
$s$.

\begin{figure}
\centering
\includegraphics[width=0.48\textwidth]{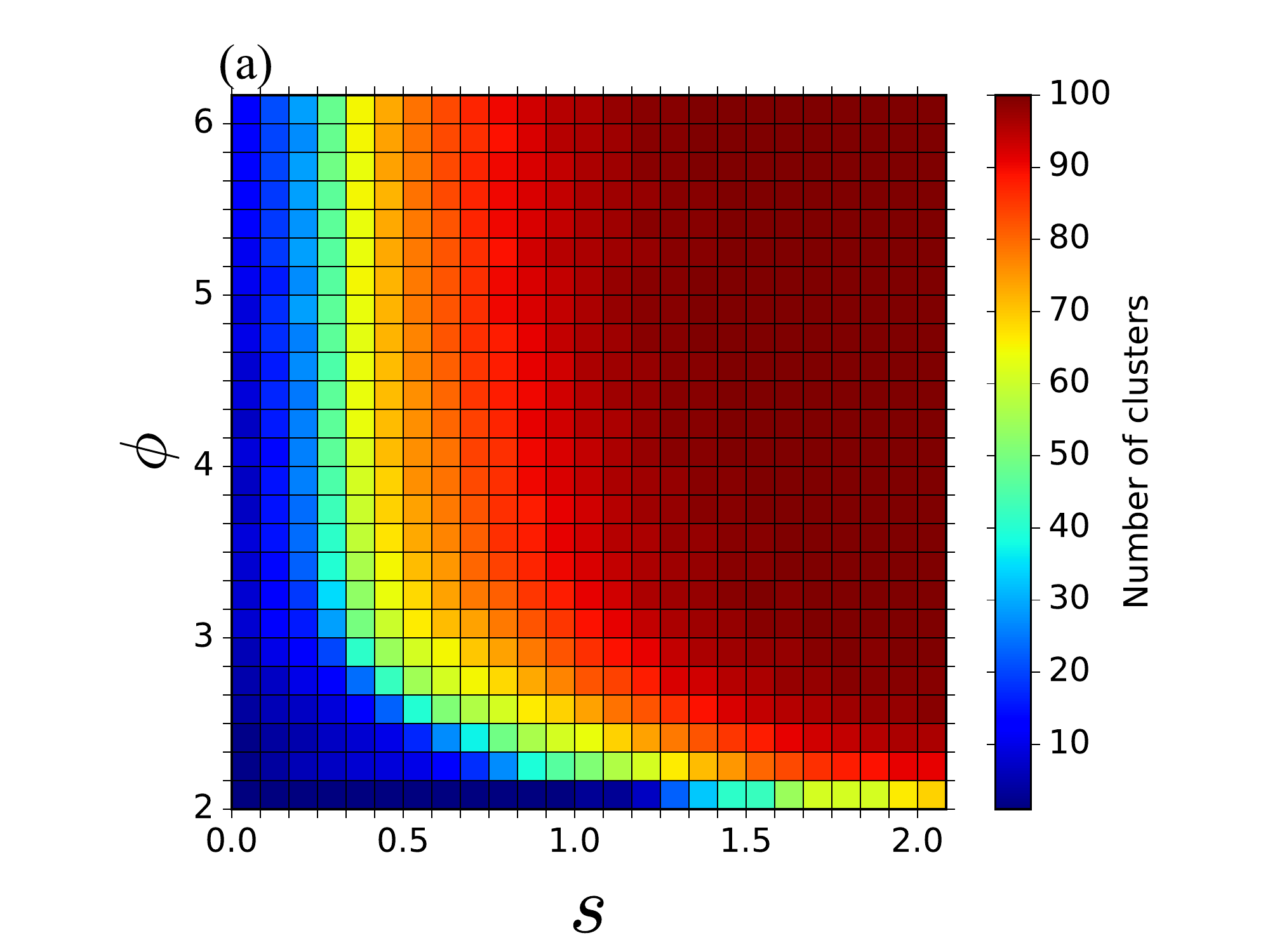}\vspace{-0.05in}
\includegraphics[width=0.48\textwidth]{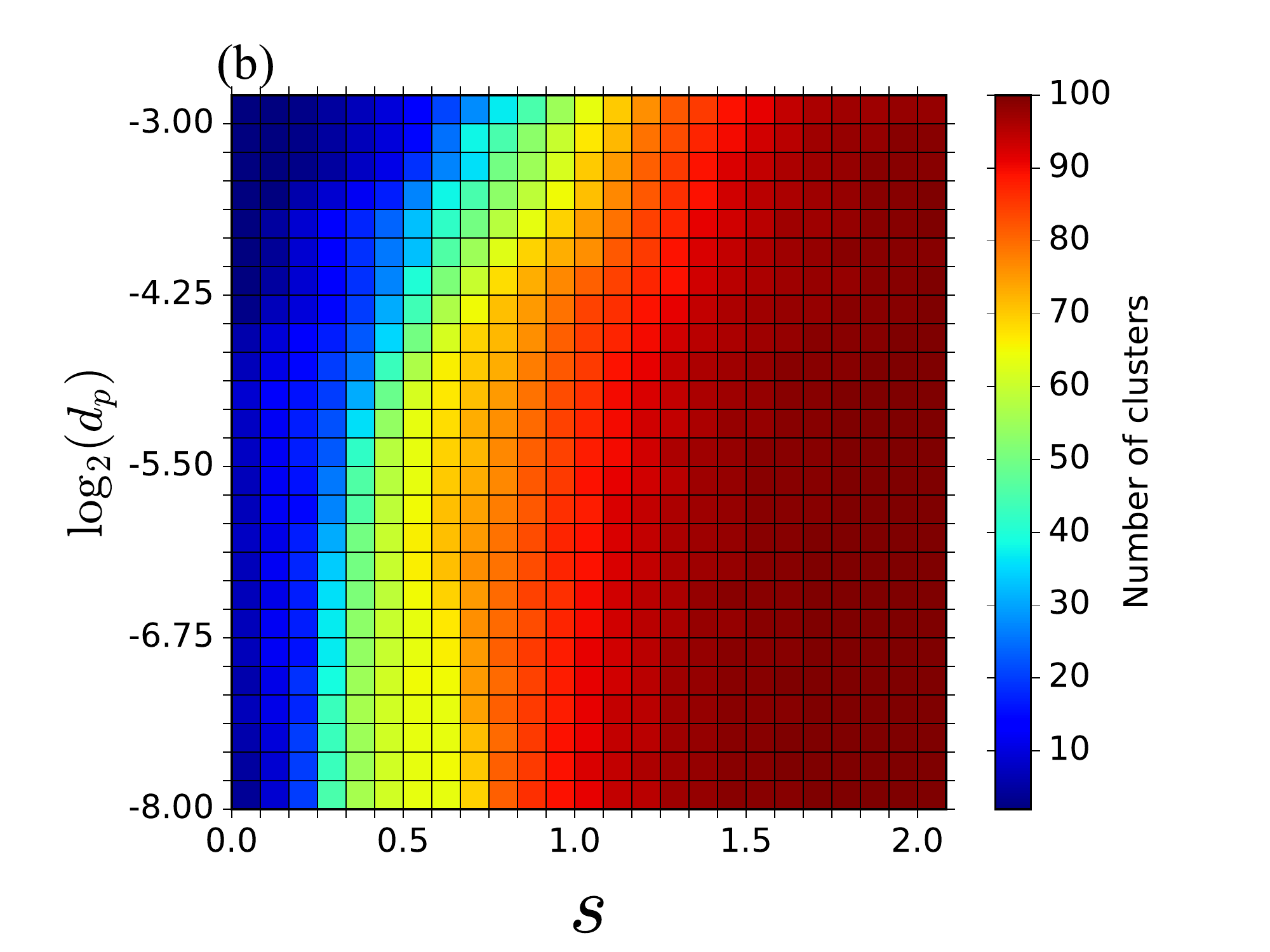}\vspace{-0.05in}
\caption{(Color online) Variation of cluster solutions with spatial
  and dispersal parameters: (a) Dispersal rates are $d_h = 2^{-5}$ and
  $d_p = 2^{-6})$ and (b) prey dispersal rate $d_h = 2^{-5}$.  Local
  dynamics are governed by weak predation $\phi = 3.0$, strong prey
  self-regulation $\theta = 0.3$ and predator mortality rate $\eta =
  1$.  Each grid summarizes 120 numerical simulations carried out
  independently.}\label{fig6}

\end{figure}

\subsection{Combined effect of system parameters and $s$ in a large network}

Until now we performed our study with an ecological network of 11
nodes. Here, we consider a relatively large network of 101 nodes and
study their dynamics. In this case, we see a variety of $k$-cluster
solutions varying from global synchrony ($k = 1$) to global asynchrony
($k = 101$). We varied the predator efficiency and predator dispersal
rate and took the mean of $k$-cluster solutions at time $t = 5000$ for
120 simulations.  Figures~\ref{fig6}(a) and \ref{fig6}(b) represent
how the cluster solutions depend on a local parameter $\phi$ and a
dispersal parameter $d_p$ for a large network of size 101 nodes.  For
a very low value of predator efficiency ($\phi = 2$) in
Fig.~\ref{fig6}(a), we see the least number of clusters as compared to
other values of $\phi$. Increasing the predator efficiency $\phi$
leads to higher number of clusters, however this is not true when $s$
is very low.  With increasing $s$, the number of clusters increases
thereby making the system asynchronous leading to higher persistence.
Surprisingly, at $s = 0$, the number of clusters is very low
irrespective of $\phi$. This tells us that when dispersal is
independent of distance, the rate of synchronization is very high and
as a consequence it may lead to global extinction.

Figure~\ref{fig6}(b) depicts how the average number of $k$-clusters
vary with predator dispersal rate ($d_p$) and power-law exponent
($s$). Higher predator dispersal leads to lower number of $k$-clusters
as compared to lower predator dispersal, which is also intuitive
because higher dispersal leads to higher synchronization in
sub-populations leading to less number of clusters. However, for higher
values of $s$ (i.e., near $s=2$), the population is asynchronised,
which is independent of the predator dispersal rate. At $s = 0$,
irrespective of the dispersal rates the number of clusters is very low
from which we can infer that even in low dispersal rate the synchrony
is very high when dispersal is independent of distance.

\subsection{Effect of network size}

\begin{figure}
\centering
\includegraphics[width=0.48\textwidth]{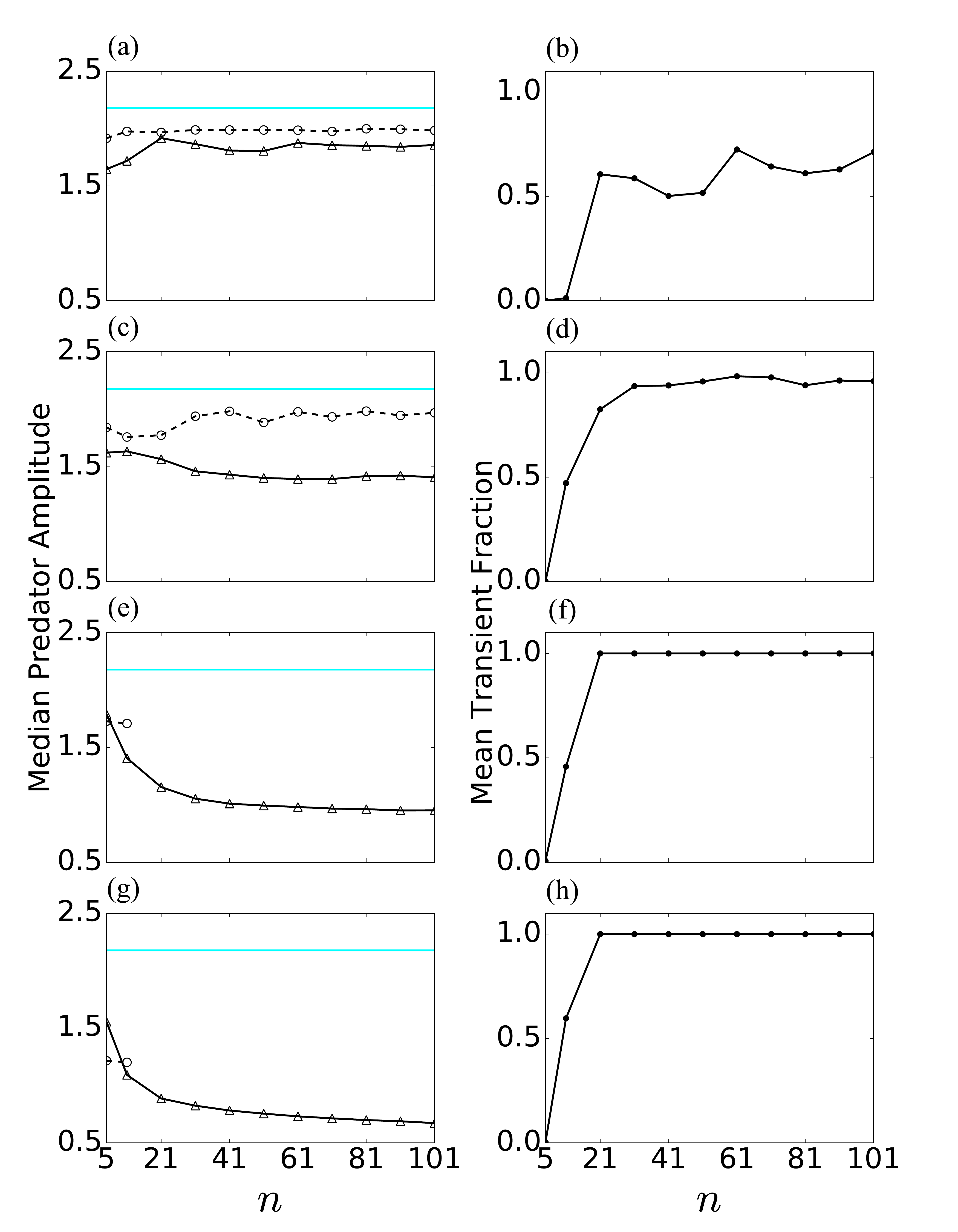}\vspace{-0.1in}
\caption{(Color online) Predator amplitude and transient duration as a
  function of network size.  Median total predator amplitude during
  transient (triangles, solid lines) and asymptotic (open circles,
  dashed lines) solution phases for a regular network. The cyan line
  corresponds to global synchronous solution for the chosen parameter
  values: $\eta = 1,\; \theta = 0.3,\; d_h = 2^{-5}$, and $d_p =
  2^{-6}$. (a,b) $\phi = 3,\; s = 0$; (c,d) $\phi = 3,\; s = 0.3$;
  (e,f) $\phi = 3,\; s = 0.5$; and (g,h) $\phi = 3,\; s = 1.0$. The
  above result corresponds to 150 numerical simulations carried out
  independently.}\label{fig7}
\end{figure}

Next, we study the influence of network size on the MPA and transient
time. Figure~\ref{fig7} illustrates how the MPA and transient time
vary with the network size for a particular $s$. It can be seen that
the transient fraction increases and MPA during transient regime
decreases with increase in the network size.  For $s > 0$, the
transient fraction saturates to one as the network size
increases. Remarkably, we find that with increasing network size the
MPA decreases thereby suggesting that smaller networks have higher MPA
and thus a lower species persistence.

The motivation behind considering variable network size in our study
is to understand the effect of metapopulation size in the dynamics of
the considered ecological network \cite{HaMoGy96:Amnat}.  We find that
small metapopulations (i.e., with less number of patches) further
increases the risk of global extinction because smaller networks are
much easier to synchronize as compared to larger networks.  Since, a
small network will have less number of $k$-cluster solution as
compared to a large network.  Hence, large metapopulations which are
spread over large landscapes (with more spatially separated patches),
more known areas and connectivities are less prone to extinction as
they are difficult to synchronize and hence carries the characteristic
of higher species persistence.

\section{Discussions and Conclusions}\label{IV}

Synchrony and stability are often considered as conflicting outcomes
of dispersal \cite{briggs2004stabilizing}. Stability here refers to
the state when populations linked by dispersal are more persistent and
hence less prone to extinction than the isolated ones. However,
synchrony on the other hand can result in global extinction.  In many
studies it has been shown that heterogeneous network is required to
generate stability and asynchrony simultaneously
\cite{holland2008strong, singh2004role,loreau2003biodiversity,
  briggs2004stabilizing}. On the contrary, we use a regular network
with power-law coupling to induce stability and asynchrony.

We have presented the importance of distance-dependent dispersal in
the ecological dynamics. We have shown how the ecological dynamics
vary when the power-law exponent $s$ is changed. For $s = 0$, the
species has equal likelihood of going to any habitat patch resulting
in high synchrony.  In other words, considering the same intensity of
both SDD and LDD can lead to a high synchronous solution which has a
high chance of global extinction. One might argue why not consider
only SDD, i.e, to allow the connection between patches that are the
nearest ones. But that would underestimate the metapopulation
effect. Because in reality, one cannot deny the fact that there are
some rare LDD that play a key role in patch recolonization. So, we
have incorporated power-law dispersal in our model with $s > 0$, as it
takes into account the fact that the intensity of dispersal reduces
with distance \cite{alex2005dispersal}.

Keeping in mind the intensity of both SDD and LDD, we have introduced
heterogeneity in the system through dispersal strength which is
distance-dependent power-law.  Using cluster analysis by calculating
correlation coefficient, we have shown that changes in the power-law
exponent $s$ can surprisingly change the ecological dynamics. As there
is a consistent increase in the number of clusters when the power-law
exponent is increased.  Larger $k$-cluster solutions have a longer
period of asynchronous dynamics and most of these solutions have
longer transients, which indicates the species persistence in
metapopulations in ecologically relevant time scales
\cite{hastings2004transients,hastings2001transient}.  There are also
several measures to quantify how good is the synchronization. In this
paper, the interpatch synchronization is quantified by using the
synchrony order parameter.  Our results indicate that with an increase
in the power-law exponent the synchrony order parameter reduces,
resulting lower synchrony/higher asynchrony with increase in $s$.

In addition, we have also shown how the MPA for transient time
decreases with increasing power-law exponent.  Hence, for larger
values of power-law exponent the species has less likelihood of
extinction during the transient phase rather than the asymptotic phase
as the amplitudes are higher at the asymptotic phases.  Moreover, we
also explore the combined effect of system parameters and power-law
exponent on the species persistence in a large network by calculating
the number of clusters.  Once we have a larger value of power-law
exponent it always supports higher asynchrony by forming more number
of clusters with variations in either local or spatial parameters.

We further demonstrated that a larger network size supports a lower
MPA in comparison with small network size.  This suggests that large
metapopulations have less risk of global extinction in comparison with
small metapopulation with few connectivities.  This finding is
significant in the context of biodiversity as larger connectivity
supports species persistence.

Future research can be initiated to observe these results
experimentally using the network of microcosms similar to the
experimental setup in \cite{holyoak1996persistence}, where it has been
experimentally shown how asynchronicity enhances species persistence
with a different network topology.

\begin{acknowledgments}
A.G. and P.S.D. acknowledge financial support from SERB, DST,
Govt. of India [Grant No.: YSS/2014/000057]. A.G. also acknowledges
fruitful discussions with Anirban Banerjee.  T.B. acknowledges
financial support from SERB, DST, Govt. of India [Grant
  No. SB/FTP/PS-005/2013].
\end{acknowledgments}


\begin{thebibliography}{10}%
\makeatletter
\providecommand \@ifxundefined [1]{%
 \ifx #1\undefined \expandafter \@firstoftwo
 \else \expandafter \@secondoftwo
\fi
}%
\providecommand \@ifnum [1]{%
 \ifnum #1\expandafter \@firstoftwo
 \else \expandafter \@secondoftwo
\fi
}%
\providecommand \enquote [1]{``#1''}%
\providecommand \bibnamefont  [1]{#1}%
\providecommand \bibfnamefont [1]{#1}%
\providecommand \citenamefont [1]{#1}%
\providecommand\href[0]{\@sanitize\@href}%
\providecommand\@href[1]{\endgroup\@@startlink{#1}\endgroup\@@href}%
\providecommand\@@href[1]{#1\@@endlink}%
\providecommand \@sanitize [0]{\begingroup\catcode`\&12\catcode`\#12\relax}%
\@ifxundefined \pdfoutput {\@firstoftwo}{%
 \@ifnum{\z@=\pdfoutput}{\@firstoftwo}{\@secondoftwo}%
}{%
 \providecommand\@@startlink[1]{\leavevmode}%
 \providecommand\@@endlink[0]{}%
}{%
 \providecommand\@@startlink[1]{%
  \leavevmode
  \pdfstartlink
   attr{/Border[0 0 1 ]/H/I/C[0 1 1]}%
   user{/Subtype/Link/A<</Type/Action/S/URI/URI(#1)>>}%
  \relax
 }%
 \providecommand\@@endlink[0]{\pdfendlink}%
}%
\providecommand \url  [0]{\begingroup\@sanitize \@url }%
\providecommand \@url [1]{\endgroup\@href {#1}{\urlprefix}}%
\providecommand \urlprefix [0]{URL }%
\providecommand \Eprint[0]{\href }%
\@ifxundefined \urlstyle {%
  \providecommand \doi [1]{doi:\discretionary{}{}{}#1}%
}{%
  \providecommand \doi [0]{doi:\discretionary{}{}{}\begingroup
  \urlstyle{rm}\Url }%
}%
\providecommand \doibase [0]{http://dx.doi.org/}%
\providecommand \Doi[1]{\href{\doibase#1}}%
\providecommand \bibAnnote [3]{%
  \BibitemShut{#1}%
  \begin{quotation}\noindent
    \textsc{Key:}\ #2\\\textsc{Annotation:}\ #3%
  \end{quotation}%
}%
\providecommand \bibAnnoteFile [2]{%
  \IfFileExists{#2}{\bibAnnote {#1} {#2} {\input{#2}}}{}%
}%
\providecommand \typeout [0]{\immediate \write \m@ne }%
\providecommand \selectlanguage [0]{\@gobble}%
\providecommand \bibinfo [0]{\@secondoftwo}%
\providecommand \bibfield [0]{\@secondoftwo}%
\providecommand \translation [1]{[#1]}%
\providecommand \BibitemOpen[0]{}%
\providecommand \bibitemStop [0]{}%
\providecommand \bibitemNoStop [0]{.\EOS\space}%
\providecommand \EOS [0]{\spacefactor3000\relax}%
\providecommand \BibitemShut [1]{\csname bibitem#1\endcsname}%
\bibitem{Han98:Nature}%
  \BibitemOpen
  \bibfield{author}{%
  \bibinfo {author} {\bibfnamefont{I.}~\bibnamefont{Hanski}},\ }%
  \bibfield{journal}{%
  \bibinfo {journal} {Nature}\ }%
  \textbf{\bibinfo {volume} {396}},\ \bibinfo {pages} {41} (\bibinfo {year}
  {1998})%
  \bibAnnoteFile{NoStop}{Han98:Nature}%
\bibitem{ilmari2003long}%
  \BibitemOpen
  \bibfield{author}{%
  \bibinfo {author} {\bibfnamefont{V.}~\bibnamefont{Ilmari~Pajunen}}\ and\
  \bibinfo {author} {\bibfnamefont{I.}~\bibnamefont{Pajunen}},\ }%
  \bibfield{journal}{%
  \bibinfo {journal} {Ecography}\ }%
  \textbf{\bibinfo {volume} {26}},\ \bibinfo {pages} {731} (\bibinfo {year}
  {2003})%
  \bibAnnoteFile{NoStop}{ilmari2003long}%
\bibitem{johst2002metapopulation}%
  \BibitemOpen
  \bibfield{author}{%
  \bibinfo {author} {\bibfnamefont{K.}~\bibnamefont{Johst}}, \bibinfo {author}
  {\bibfnamefont{R.}~\bibnamefont{Brandl}},\ and\ \bibinfo {author}
  {\bibfnamefont{S.}~\bibnamefont{Eber}},\ }%
  \bibfield{journal}{%
  \bibinfo {journal} {Oikos}\ }%
  \textbf{\bibinfo {volume} {98}},\ \bibinfo {pages} {263} (\bibinfo {year}
  {2002})%
  \bibAnnoteFile{NoStop}{johst2002metapopulation}%
\bibitem{gyllenberg1992single}%
  \BibitemOpen
  \bibfield{author}{%
  \bibinfo {author} {\bibfnamefont{M.}~\bibnamefont{Gyllenberg}}\ and\ \bibinfo
  {author} {\bibfnamefont{I.}~\bibnamefont{Hanski}},\ }%
  \bibfield{journal}{%
  \bibinfo {journal} {Theoretical Population Biology}\ }%
  \textbf{\bibinfo {volume} {42}},\ \bibinfo {pages} {35} (\bibinfo {year}
  {1992})%
  \bibAnnoteFile{NoStop}{gyllenberg1992single}%
\bibitem{roy2005temporal}%
  \BibitemOpen
  \bibfield{author}{%
  \bibinfo {author} {\bibfnamefont{M.}~\bibnamefont{Roy}}, \bibinfo {author}
  {\bibfnamefont{R.~D.}\ \bibnamefont{Holt}},\ and\ \bibinfo {author}
  {\bibfnamefont{M.}~\bibnamefont{Barfield}},\ }%
  \bibfield{journal}{%
  \bibinfo {journal} {The American Naturalist}\ }%
  \textbf{\bibinfo {volume} {166}},\ \bibinfo {pages} {246} (\bibinfo {year}
  {2005})%
  \bibAnnoteFile{NoStop}{roy2005temporal}%
\bibitem{roy2008generalizing}%
  \BibitemOpen
  \bibfield{author}{%
  \bibinfo {author} {\bibfnamefont{M.}~\bibnamefont{Roy}}, \bibinfo {author}
  {\bibfnamefont{K.}~\bibnamefont{Harding}},\ and\ \bibinfo {author}
  {\bibfnamefont{R.~D.}\ \bibnamefont{Holt}},\ }%
  \bibfield{journal}{%
  \bibinfo {journal} {Journal of Theoretical Biology}\ }%
  \textbf{\bibinfo {volume} {255}},\ \bibinfo {pages} {152} (\bibinfo {year}
  {2008})%
  \bibAnnoteFile{NoStop}{roy2008generalizing}%
\bibitem{mccallum2002disease}%
  \BibitemOpen
  \bibfield{author}{%
  \bibinfo {author} {\bibfnamefont{H.}~\bibnamefont{McCallum}}\ and\ \bibinfo
  {author} {\bibfnamefont{A.}~\bibnamefont{Dobson}},\ }%
  \bibfield{journal}{%
  \bibinfo {journal} {Proceedings of the Royal Society of London B: Biological
  Sciences}\ }%
  \textbf{\bibinfo {volume} {269}},\ \bibinfo {pages} {2041} (\bibinfo {year}
  {2002})%
  \bibAnnoteFile{NoStop}{mccallum2002disease}%
\bibitem{sultan2002metapopulation}%
  \BibitemOpen
  \bibfield{author}{%
  \bibinfo {author} {\bibfnamefont{S.~E.}\ \bibnamefont{Sultan}}\ and\ \bibinfo
  {author} {\bibfnamefont{H.~G.}\ \bibnamefont{Spencer}},\ }%
  \bibfield{journal}{%
  \bibinfo {journal} {The American Naturalist}\ }%
  \textbf{\bibinfo {volume} {160}},\ \bibinfo {pages} {271} (\bibinfo {year}
  {2002})%
  \bibAnnoteFile{NoStop}{sultan2002metapopulation}%
\bibitem{joshi2001local}%
  \BibitemOpen
  \bibfield{author}{%
  \bibinfo {author} {\bibfnamefont{J.}~\bibnamefont{Joshi}}, \bibinfo {author}
  {\bibfnamefont{B.}~\bibnamefont{Schmid}}, \bibinfo {author}
  {\bibfnamefont{M.}~\bibnamefont{Caldeira}}, \bibinfo {author}
  {\bibfnamefont{P.}~\bibnamefont{Dimitrakopoulos}}, \bibinfo {author}
  {\bibfnamefont{J.}~\bibnamefont{Good}}, \bibinfo {author}
  {\bibfnamefont{R.}~\bibnamefont{Harris}}, \bibinfo {author}
  {\bibfnamefont{A.}~\bibnamefont{Hector}}, \bibinfo {author}
  {\bibfnamefont{K.}~\bibnamefont{Huss-Danell}}, \bibinfo {author}
  {\bibfnamefont{A.}~\bibnamefont{Jumpponen}}, \bibinfo {author}
  {\bibfnamefont{A.}~\bibnamefont{Minns}}, \emph{et~al.},\ }%
  \bibfield{journal}{%
  \bibinfo {journal} {Ecology Letters}\ }%
  \textbf{\bibinfo {volume} {4}},\ \bibinfo {pages} {536} (\bibinfo {year}
  {2001})%
  \bibAnnoteFile{NoStop}{joshi2001local}%
\bibitem{hansson1991dispersal}%
  \BibitemOpen
  \bibfield{author}{%
  \bibinfo {author} {\bibfnamefont{L.}~\bibnamefont{Hansson}},\ }%
  \bibfield{journal}{%
  \bibinfo {journal} {Biological journal of the Linnean Society}\ }%
  \textbf{\bibinfo {volume} {42}},\ \bibinfo {pages} {89} (\bibinfo {year}
  {1991})%
  \bibAnnoteFile{NoStop}{hansson1991dispersal}%
\bibitem{holyoak1996persistence}%
  \BibitemOpen
  \bibfield{author}{%
  \bibinfo {author} {\bibfnamefont{M.}~\bibnamefont{Holyoak}}\ and\ \bibinfo
  {author} {\bibfnamefont{S.~P.}\ \bibnamefont{Lawler}},\ }%
  \bibfield{journal}{%
  \bibinfo {journal} {Ecology}\ }%
  \textbf{\bibinfo {volume} {77}},\ \bibinfo {pages} {1867} (\bibinfo {year}
  {1996})%
  \bibAnnoteFile{NoStop}{holyoak1996persistence}%
\bibitem{heino1997synchronous}%
  \BibitemOpen
  \bibfield{author}{%
  \bibinfo {author} {\bibfnamefont{M.}~\bibnamefont{Heino}}, \bibinfo {author}
  {\bibfnamefont{V.}~\bibnamefont{Kaitala}}, \bibinfo {author}
  {\bibfnamefont{E.}~\bibnamefont{Ranta}},\ and\ \bibinfo {author}
  {\bibfnamefont{J.}~\bibnamefont{Lindstr{\"o}m}},\ }%
  \bibfield{journal}{%
  \bibinfo {journal} {Proceedings of the Royal Society of London B: Biological
  Sciences}\ }%
  \textbf{\bibinfo {volume} {264}},\ \bibinfo {pages} {481} (\bibinfo {year}
  {1997})%
  \bibAnnoteFile{NoStop}{heino1997synchronous}%
\bibitem{hudson1999moran}%
  \BibitemOpen
  \bibfield{author}{%
  \bibinfo {author} {\bibfnamefont{P.~J.}\ \bibnamefont{Hudson}}\ and\ \bibinfo
  {author} {\bibfnamefont{I.~M.}\ \bibnamefont{Cattadori}},\ }%
  \bibfield{journal}{%
  \bibinfo {journal} {Trends in Ecology \& Evolution}\ }%
  \textbf{\bibinfo {volume} {14}},\ \bibinfo {pages} {1} (\bibinfo {year}
  {1999})%
  \bibAnnoteFile{NoStop}{hudson1999moran}%
\bibitem{hanski1995metapopulation}%
  \BibitemOpen
  \bibfield{author}{%
  \bibinfo {author} {\bibfnamefont{I.}~\bibnamefont{Hanski}}, \bibinfo {author}
  {\bibfnamefont{T.}~\bibnamefont{Pakkala}}, \bibinfo {author}
  {\bibfnamefont{M.}~\bibnamefont{Kuussaari}},\ and\ \bibinfo {author}
  {\bibfnamefont{G.}~\bibnamefont{Lei}},\ }%
  \bibfield{journal}{%
  \bibinfo {journal} {Oikos},\ \bibinfo {pages} {21}}%
   (\bibinfo {year} {1995})%
  \bibAnnoteFile{NoStop}{hanski1995metapopulation}%
\bibitem{ellner2001habitat}%
  \BibitemOpen
  \bibfield{author}{%
  \bibinfo {author} {\bibfnamefont{S.~P.}\ \bibnamefont{Ellner}}, \bibinfo
  {author} {\bibfnamefont{E.}~\bibnamefont{McCauley}}, \bibinfo {author}
  {\bibfnamefont{B.~E.}\ \bibnamefont{Kendall}}, \bibinfo {author}
  {\bibfnamefont{C.~J.}\ \bibnamefont{Briggs}}, \bibinfo {author}
  {\bibfnamefont{P.~R.}\ \bibnamefont{Hosseini}}, \bibinfo {author}
  {\bibfnamefont{S.~N.}\ \bibnamefont{Wood}}, \bibinfo {author}
  {\bibfnamefont{A.}~\bibnamefont{Janssen}}, \bibinfo {author}
  {\bibfnamefont{M.~W.}\ \bibnamefont{Sabelis}}, \bibinfo {author}
  {\bibfnamefont{P.}~\bibnamefont{Turchin}}, \bibinfo {author}
  {\bibfnamefont{R.~M.}\ \bibnamefont{Nisbet}}, \emph{et~al.},\ }%
  \bibfield{journal}{%
  \bibinfo {journal} {Nature}\ }%
  \textbf{\bibinfo {volume} {412}},\ \bibinfo {pages} {538} (\bibinfo {year}
  {2001})%
  \bibAnnoteFile{NoStop}{ellner2001habitat}%
\bibitem{dey2006stability}%
  \BibitemOpen
  \bibfield{author}{%
  \bibinfo {author} {\bibfnamefont{S.}~\bibnamefont{Dey}}\ and\ \bibinfo
  {author} {\bibfnamefont{A.}~\bibnamefont{Joshi}},\ }%
  \bibfield{journal}{%
  \bibinfo {journal} {Science}\ }%
  \textbf{\bibinfo {volume} {312}},\ \bibinfo {pages} {434} (\bibinfo {year}
  {2006})%
  \bibAnnoteFile{NoStop}{dey2006stability}%
\bibitem{ringsby2002asynchronous}%
  \BibitemOpen
  \bibfield{author}{%
  \bibinfo {author} {\bibfnamefont{T.~H.}\ \bibnamefont{Ringsby}}, \bibinfo
  {author} {\bibfnamefont{B.-E.}\ \bibnamefont{Saether}}, \bibinfo {author}
  {\bibfnamefont{J.}~\bibnamefont{Tufto}}, \bibinfo {author}
  {\bibfnamefont{H.}~\bibnamefont{Jensen}},\ and\ \bibinfo {author}
  {\bibfnamefont{E.~J.}\ \bibnamefont{Solberg}},\ }%
  \bibfield{journal}{%
  \bibinfo {journal} {Ecology}\ }%
  \textbf{\bibinfo {volume} {83}},\ \bibinfo {pages} {561} (\bibinfo {year}
  {2002})%
  \bibAnnoteFile{NoStop}{ringsby2002asynchronous}%
\bibitem{holland2008strong}%
  \BibitemOpen
  \bibfield{author}{%
  \bibinfo {author} {\bibfnamefont{M.~D.}\ \bibnamefont{Holland}}\ and\
  \bibinfo {author} {\bibfnamefont{A.}~\bibnamefont{Hastings}},\ }%
  \bibfield{journal}{%
  \bibinfo {journal} {Nature}\ }%
  \textbf{\bibinfo {volume} {456}},\ \bibinfo {pages} {792} (\bibinfo {year}
  {2008})%
  \bibAnnoteFile{NoStop}{holland2008strong}%
\bibitem{BoBe09:Oikos}%
  \BibitemOpen
  \bibfield{author}{%
  \bibinfo {author} {\bibfnamefont{D.~E.}\ \bibnamefont{Bowler}}\ and\ \bibinfo
  {author} {\bibfnamefont{T.~G.}\ \bibnamefont{Benton}},\ }%
  \bibfield{journal}{%
  \bibinfo {journal} {Oikos}\ }%
  \textbf{\bibinfo {volume} {118}},\ \bibinfo {pages} {403} (\bibinfo {year}
  {2009})%
  \bibAnnoteFile{NoStop}{BoBe09:Oikos}%
\bibitem{alex2005dispersal}%
  \BibitemOpen
  \bibfield{author}{%
  \bibinfo {author} {\bibfnamefont{M.}~\bibnamefont{Alex~Smith}}\ and\ \bibinfo
  {author} {\bibfnamefont{D.}~\bibnamefont{M~Green}},\ }%
  \bibfield{journal}{%
  \bibinfo {journal} {Ecography}\ }%
  \textbf{\bibinfo {volume} {28}},\ \bibinfo {pages} {110} (\bibinfo {year}
  {2005})%
  \bibAnnoteFile{NoStop}{alex2005dispersal}%
\bibitem{baguette2003long}%
  \BibitemOpen
  \bibfield{author}{%
  \bibinfo {author} {\bibfnamefont{M.}~\bibnamefont{Baguette}},\ }%
  \bibfield{journal}{%
  \bibinfo {journal} {Ecography}\ }%
  \textbf{\bibinfo {volume} {26}},\ \bibinfo {pages} {153} (\bibinfo {year}
  {2003})%
  \bibAnnoteFile{NoStop}{baguette2003long}%
\bibitem{bohrer2005effects}%
  \BibitemOpen
  \bibfield{author}{%
  \bibinfo {author} {\bibfnamefont{G.}~\bibnamefont{Bohrer}}, \bibinfo {author}
  {\bibfnamefont{R.}~\bibnamefont{Nathan}},\ and\ \bibinfo {author}
  {\bibfnamefont{S.}~\bibnamefont{Volis}},\ }%
  \bibfield{journal}{%
  \bibinfo {journal} {Journal of Ecology}\ }%
  \textbf{\bibinfo {volume} {93}},\ \bibinfo {pages} {1029} (\bibinfo {year}
  {2005})%
  \bibAnnoteFile{NoStop}{bohrer2005effects}%
\bibitem{trakhtenbrot2005importance}%
  \BibitemOpen
  \bibfield{author}{%
  \bibinfo {author} {\bibfnamefont{A.}~\bibnamefont{Trakhtenbrot}}, \bibinfo
  {author} {\bibfnamefont{R.}~\bibnamefont{Nathan}}, \bibinfo {author}
  {\bibfnamefont{G.}~\bibnamefont{Perry}},\ and\ \bibinfo {author}
  {\bibfnamefont{D.~M.}\ \bibnamefont{Richardson}},\ }%
  \bibfield{journal}{%
  \bibinfo {journal} {Diversity and Distributions}\ }%
  \textbf{\bibinfo {volume} {11}},\ \bibinfo {pages} {173} (\bibinfo {year}
  {2005})%
  \bibAnnoteFile{NoStop}{trakhtenbrot2005importance}%
\bibitem{hanski1999metapopulation}%
  \BibitemOpen
  \bibfield{author}{%
  \bibinfo {author} {\bibfnamefont{I.}~\bibnamefont{Hanski}},\ }%
  \emph{\bibinfo {title} {Metapopulation Ecology}}\ (\bibinfo {publisher}
  {Oxford University Press},\ \bibinfo {year} {1999})%
  \bibAnnoteFile{NoStop}{hanski1999metapopulation}%
\bibitem{hill1996effects}%
  \BibitemOpen
  \bibfield{author}{%
  \bibinfo {author} {\bibfnamefont{J.}~\bibnamefont{Hill}}, \bibinfo {author}
  {\bibfnamefont{C.}~\bibnamefont{Thomas}},\ and\ \bibinfo {author}
  {\bibfnamefont{O.}~\bibnamefont{Lewis}},\ }%
  \bibfield{journal}{%
  \bibinfo {journal} {Journal of Animal Ecology},\ \bibinfo {pages} {725}}%
   (\bibinfo {year} {1996})%
  \bibAnnoteFile{NoStop}{hill1996effects}%
\bibitem{thomas1997butterfly}%
  \BibitemOpen
  \bibfield{author}{%
  \bibinfo {author} {\bibfnamefont{C.~D.}\ \bibnamefont{Thomas}}\ and\ \bibinfo
  {author} {\bibfnamefont{I.}~\bibnamefont{Hanski}},\ }%
  in\ \emph{\bibinfo {booktitle} {Metapopulation Biology: Ecology, Genetics and
  Evolution}},\ \bibinfo {editor} {edited by\ \bibinfo {editor}
  {\bibfnamefont{I.}~\bibnamefont{Hanski}}\ and\ \bibinfo {editor}
  {\bibfnamefont{M.}~\bibnamefont{Gilpin}}}\ (\bibinfo {publisher} {Academic
  Press},\ \bibinfo {year} {1997})\ pp.\ \bibinfo {pages} {359--386}%
  \bibAnnoteFile{NoStop}{thomas1997butterfly}%
\bibitem{baguette2000population}%
  \BibitemOpen
  \bibfield{author}{%
  \bibinfo {author} {\bibfnamefont{M.}~\bibnamefont{Baguette}}, \bibinfo
  {author} {\bibfnamefont{S.}~\bibnamefont{Petit}},\ and\ \bibinfo {author}
  {\bibfnamefont{F.}~\bibnamefont{Qu{\'e}va}},\ }%
  \bibfield{journal}{%
  \bibinfo {journal} {Journal of Applied Ecology}\ }%
  \textbf{\bibinfo {volume} {37}},\ \bibinfo {pages} {100} (\bibinfo {year}
  {2000})%
  \bibAnnoteFile{NoStop}{baguette2000population}%
\bibitem{PhysRevB.54.R12661}%
  \BibitemOpen
  \bibfield{author}{%
  \bibinfo {author} {\bibfnamefont{S.~A.}\ \bibnamefont{Cannas}}\ and\ \bibinfo
  {author} {\bibfnamefont{F.~A.}\ \bibnamefont{Tamarit}},\ }%
  \bibfield{journal}{%
  \bibinfo {journal} {Phys. Rev. B}\ }%
  \textbf{\bibinfo {volume} {54}},\ \bibinfo {pages} {R12661} (\bibinfo {month}
  {Nov}\ \bibinfo {year} {1996})%
  \bibAnnoteFile{NoStop}{PhysRevB.54.R12661}%
\bibitem{PhysRevLett.106.058104}%
  \BibitemOpen
  \bibfield{author}{%
  \bibinfo {author} {\bibfnamefont{N.}~\bibnamefont{Uchida}}\ and\ \bibinfo
  {author} {\bibfnamefont{R.}~\bibnamefont{Golestanian}},\ }%
  \bibfield{journal}{%
  \bibinfo {journal} {Phys. Rev. Lett.}\ }%
  \textbf{\bibinfo {volume} {106}},\ \bibinfo {pages} {058104} (\bibinfo
  {month} {Feb}\ \bibinfo {year} {2011})%
  \bibAnnoteFile{NoStop}{PhysRevLett.106.058104}%
\bibitem{C0SM01121E}%
  \BibitemOpen
  \bibfield{author}{%
  \bibinfo {author} {\bibfnamefont{R.}~\bibnamefont{Golestanian}}, \bibinfo
  {author} {\bibfnamefont{J.~M.}\ \bibnamefont{Yeomans}},\ and\ \bibinfo
  {author} {\bibfnamefont{N.}~\bibnamefont{Uchida}},\ }%
  \bibfield{journal}{%
  \bibinfo {journal} {Soft Matter}\ }%
  \textbf{\bibinfo {volume} {7}},\ \bibinfo {pages} {3074} (\bibinfo {year}
  {2011})%
  \bibAnnoteFile{NoStop}{C0SM01121E}%
\bibitem{PhysRevE.94.032206}%
  \BibitemOpen
  \bibfield{author}{%
  \bibinfo {author} {\bibfnamefont{T.}~\bibnamefont{Banerjee}}, \bibinfo
  {author} {\bibfnamefont{P.~S.}\ \bibnamefont{Dutta}}, \bibinfo {author}
  {\bibfnamefont{A.}~\bibnamefont{Zakharova}},\ and\ \bibinfo {author}
  {\bibfnamefont{E.}~\bibnamefont{Sch\"oll}},\ }%
  \bibfield{journal}{%
  \bibinfo {journal} {Phys. Rev. E}\ }%
  \textbf{\bibinfo {volume} {94}},\ \bibinfo {pages} {032206} (\bibinfo {month}
  {Sep}\ \bibinfo {year} {2016})%
  \bibAnnoteFile{NoStop}{PhysRevE.94.032206}%
\bibitem{PhysRevLett.74.3297}%
  \BibitemOpen
  \bibfield{author}{%
  \bibinfo {author} {\bibfnamefont{S.}~\bibnamefont{Raghavachari}}\ and\
  \bibinfo {author} {\bibfnamefont{J.~A.}\ \bibnamefont{Glazier}},\ }%
  \bibfield{journal}{%
  \bibinfo {journal} {Phys. Rev. Lett.}\ }%
  \textbf{\bibinfo {volume} {74}},\ \bibinfo {pages} {3297} (\bibinfo {month}
  {Apr}\ \bibinfo {year} {1995})%
  \bibAnnoteFile{NoStop}{PhysRevLett.74.3297}%
\bibitem{RoMa63}%
  \BibitemOpen
  \bibfield{author}{%
  \bibinfo {author} {\bibfnamefont{M.~L.}\ \bibnamefont{Rosenzweig}}\ and\
  \bibinfo {author} {\bibfnamefont{R.~H.}\ \bibnamefont{MacArthur}},\ }%
  \bibfield{journal}{%
  \bibinfo {journal} {The American Naturalist}\ }%
  \textbf{\bibinfo {volume} {97}},\ \bibinfo {pages} {209} (\bibinfo {year}
  {1963})%
  \bibAnnoteFile{NoStop}{RoMa63}%
\bibitem{GoHa08}%
  \BibitemOpen
  \bibfield{author}{%
  \bibinfo {author} {\bibfnamefont{E.~E.}\ \bibnamefont{Goldwyn}}\ and\
  \bibinfo {author} {\bibfnamefont{A.}~\bibnamefont{Hastings}},\ }%
  \bibfield{journal}{%
  \bibinfo {journal} {Theoretical Population Biology}\ }%
  \textbf{\bibinfo {volume} {73}},\ \bibinfo {pages} {395} (\bibinfo {year}
  {2008})%
  \bibAnnoteFile{NoStop}{GoHa08}%
\bibitem{GoHa11}%
  \BibitemOpen
  \bibfield{author}{%
  \bibinfo {author} {\bibfnamefont{E.~E.}\ \bibnamefont{Goldwyn}}\ and\
  \bibinfo {author} {\bibfnamefont{A.}~\bibnamefont{Hastings}},\ }%
  \bibfield{journal}{%
  \bibinfo {journal} {Journal of Theoretical Biology}\ }%
  \textbf{\bibinfo {volume} {289}},\ \bibinfo {pages} {237} (\bibinfo {year}
  {2011})%
  \bibAnnoteFile{NoStop}{GoHa11}%
\bibitem{hastings2001transient}%
  \BibitemOpen
  \bibfield{author}{%
  \bibinfo {author} {\bibfnamefont{A.}~\bibnamefont{Hastings}},\ }%
  \bibfield{journal}{%
  \bibinfo {journal} {Ecology Letters}\ }%
  \textbf{\bibinfo {volume} {4}},\ \bibinfo {pages} {215} (\bibinfo {year}
  {2001})%
  \bibAnnoteFile{NoStop}{hastings2001transient}%
\bibitem{wall2013synchronization}%
  \BibitemOpen
  \bibfield{author}{%
  \bibinfo {author} {\bibfnamefont{E.}~\bibnamefont{Wall}}, \bibinfo {author}
  {\bibfnamefont{F.}~\bibnamefont{Guichard}},\ and\ \bibinfo {author}
  {\bibfnamefont{A.~R.}\ \bibnamefont{Humphries}},\ }%
  \bibfield{journal}{%
  \bibinfo {journal} {Theoretical Ecology}\ }%
  \textbf{\bibinfo {volume} {6}},\ \bibinfo {pages} {405} (\bibinfo {year}
  {2013})%
  \bibAnnoteFile{NoStop}{wall2013synchronization}%
\bibitem{cohen1996cvode}%
  \BibitemOpen
  \bibfield{author}{%
  \bibinfo {author} {\bibfnamefont{S.~D.}\ \bibnamefont{Cohen}}\ and\ \bibinfo
  {author} {\bibfnamefont{A.~C.}\ \bibnamefont{Hindmarsh}},\ }%
  \bibfield{journal}{%
  \bibinfo {journal} {Computers in Physics}\ }%
  \textbf{\bibinfo {volume} {10}},\ \bibinfo {pages} {138} (\bibinfo {year}
  {1996})%
  \bibAnnoteFile{NoStop}{cohen1996cvode}%
\bibitem{komin2011synchronization}%
  \BibitemOpen
  \bibfield{author}{%
  \bibinfo {author} {\bibfnamefont{N.}~\bibnamefont{Komin}}, \bibinfo {author}
  {\bibfnamefont{A.~C.}\ \bibnamefont{Murza}}, \bibinfo {author}
  {\bibfnamefont{E.}~\bibnamefont{Hern{\'a}ndez-Garc{\'\i}a}},\ and\ \bibinfo
  {author} {\bibfnamefont{R.}~\bibnamefont{Toral}},\ }%
  \bibfield{journal}{%
  \bibinfo {journal} {Interface Focus}\ }%
  \textbf{\bibinfo {volume} {1}},\ \bibinfo {pages} {167} (\bibinfo {year}
  {2011})%
  \bibAnnoteFile{NoStop}{komin2011synchronization}%
\bibitem{hastings2004transients}%
  \BibitemOpen
  \bibfield{author}{%
  \bibinfo {author} {\bibfnamefont{A.}~\bibnamefont{Hastings}},\ }%
  \bibfield{journal}{%
  \bibinfo {journal} {Trends in Ecology \& Evolution}\ }%
  \textbf{\bibinfo {volume} {19}},\ \bibinfo {pages} {39} (\bibinfo {year}
  {2004})%
  \bibAnnoteFile{NoStop}{hastings2004transients}%
\bibitem{huisman1999biodiversity}%
  \BibitemOpen
  \bibfield{author}{%
  \bibinfo {author} {\bibfnamefont{J.}~\bibnamefont{Huisman}}\ and\ \bibinfo
  {author} {\bibfnamefont{F.~J.}\ \bibnamefont{Weissing}},\ }%
  \bibfield{journal}{%
  \bibinfo {journal} {Nature}\ }%
  \textbf{\bibinfo {volume} {402}},\ \bibinfo {pages} {407} (\bibinfo {year}
  {1999})%
  \bibAnnoteFile{NoStop}{huisman1999biodiversity}%
\bibitem{HaMoGy96:Amnat}%
  \BibitemOpen
  \bibfield{author}{%
  \bibinfo {author} {\bibfnamefont{I.}~\bibnamefont{Hanski}}, \bibinfo {author}
  {\bibfnamefont{A.}~\bibnamefont{Moilanen}},\ and\ \bibinfo {author}
  {\bibfnamefont{M.}~\bibnamefont{Gyllenberg}},\ }%
  \bibfield{journal}{%
  \bibinfo {journal} {The American Naturalist}\ }%
  \textbf{\bibinfo {volume} {147}},\ \bibinfo {pages} {527} (\bibinfo {year}
  {1996})%
  \bibAnnoteFile{NoStop}{HaMoGy96:Amnat}%
\bibitem{briggs2004stabilizing}%
  \BibitemOpen
  \bibfield{author}{%
  \bibinfo {author} {\bibfnamefont{C.~J.}\ \bibnamefont{Briggs}}\ and\ \bibinfo
  {author} {\bibfnamefont{M.~F.}\ \bibnamefont{Hoopes}},\ }%
  \bibfield{journal}{%
  \bibinfo {journal} {Theoretical Population Biology}\ }%
  \textbf{\bibinfo {volume} {65}},\ \bibinfo {pages} {299} (\bibinfo {year}
  {2004})%
  \bibAnnoteFile{NoStop}{briggs2004stabilizing}%
\bibitem{singh2004role}%
  \BibitemOpen
  \bibfield{author}{%
  \bibinfo {author} {\bibfnamefont{B.~K.}\ \bibnamefont{Singh}}, \bibinfo
  {author} {\bibfnamefont{J.~S.}\ \bibnamefont{Rao}}, \bibinfo {author}
  {\bibfnamefont{R.}~\bibnamefont{Ramaswamy}},\ and\ \bibinfo {author}
  {\bibfnamefont{S.}~\bibnamefont{Sinha}},\ }%
  \bibfield{journal}{%
  \bibinfo {journal} {Ecological Modelling}\ }%
  \textbf{\bibinfo {volume} {180}},\ \bibinfo {pages} {435} (\bibinfo {year}
  {2004})%
  \bibAnnoteFile{NoStop}{singh2004role}%
\bibitem{loreau2003biodiversity}%
  \BibitemOpen
  \bibfield{author}{%
  \bibinfo {author} {\bibfnamefont{M.}~\bibnamefont{Loreau}}, \bibinfo {author}
  {\bibfnamefont{N.}~\bibnamefont{Mouquet}},\ and\ \bibinfo {author}
  {\bibfnamefont{A.}~\bibnamefont{Gonzalez}},\ }%
  \bibfield{journal}{%
  \bibinfo {journal} {Proceedings of the National Academy of Sciences}\ }%
  \textbf{\bibinfo {volume} {100}},\ \bibinfo {pages} {12765} (\bibinfo {year}
  {2003})%
  \bibAnnoteFile{NoStop}{loreau2003biodiversity}%
\end{thebibliography}

\providecommand{\noopsort}[1]{}\providecommand{\singleletter}[1]{#1}%

\end{document}